\documentclass[twocolumn]{aastex631}

\usepackage{amsmath,amsfonts,amssymb}
\usepackage{graphicx}
\usepackage{enumitem}
\usepackage{gensymb}

\begin{document}

\title{The MAPS Adaptive Secondary Mirror: \\ First Light, Laboratory Work, and Achievements}

\author{Jess A. Johnson}
 \affiliation{Steward Observatory, University of Arizona, 933 N. Cherry Ave., Tucson AZ 85721, USA}
\author{Amali Vaz}
 \affiliation{Steward Observatory, University of Arizona, 933 N. Cherry Ave., Tucson AZ 85721, USA}
\author{Manny Montoya}
 \affiliation{Steward Observatory, University of Arizona, 933 N. Cherry Ave., Tucson AZ 85721, USA}
\author{Narsireddy Anugu}
 \affiliation{Steward Observatory, University of Arizona, 933 N. Cherry Ave., Tucson AZ 85721, USA}
\author{Cameron Ard}
 \affiliation{Steward Observatory, University of Arizona, 933 N. Cherry Ave., Tucson AZ 85721, USA}
\author{Jared Carlson}
 \affiliation{Steward Observatory, University of Arizona, 933 N. Cherry Ave., Tucson AZ 85721, USA}
\author{Kimberly Chapman}
 \affiliation{Steward Observatory, University of Arizona, 933 N. Cherry Ave., Tucson AZ 85721, USA}
\author{Olivier Durney}
 \affiliation{Steward Observatory, University of Arizona, 933 N. Cherry Ave., Tucson AZ 85721, USA}
\author{Chuck Fellows}
 \affiliation{Steward Observatory, University of Arizona, 933 N. Cherry Ave., Tucson AZ 85721, USA}
\author{Andrew Gardner}
 \affiliation{Steward Observatory, University of Arizona, 933 N. Cherry Ave., Tucson AZ 85721, USA}
\author{Olivier Guyon}
 \affiliation{Steward Observatory, University of Arizona, 933 N. Cherry Ave., Tucson AZ 85721, USA}
\author{Buell Jannuzi}
 \affiliation{Steward Observatory, University of Arizona, 933 N. Cherry Ave., Tucson AZ 85721, USA}
\author{Ron Jones}
 \affiliation{Steward Observatory, University of Arizona, 933 N. Cherry Ave., Tucson AZ 85721, USA}
\author{Craig Kulesa}
 \affiliation{Steward Observatory, University of Arizona, 933 N. Cherry Ave., Tucson AZ 85721, USA}
\author{Joseph Long}
 \affiliation{Steward Observatory, University of Arizona, 933 N. Cherry Ave., Tucson AZ 85721, USA}
\author{Eden McEwen}
 \affiliation{Steward Observatory, University of Arizona, 933 N. Cherry Ave., Tucson AZ 85721, USA}
\author{Jared Males}
 \affiliation{Steward Observatory, University of Arizona, 933 N. Cherry Ave., Tucson AZ 85721, USA}
\author{Emily Mailhot}
 \affiliation{Steward Observatory, University of Arizona, 933 N. Cherry Ave., Tucson AZ 85721, USA}
\author{Jorge Sanchez}
 \affiliation{School of Earth and Space Exploration, Arizona State University, Tempe, AZ 85281, USA}
\author{Suresh Sivanandam}
 \affiliation{David A. Dunlap Department of Astronomy \& Astrophysics, University of Toronto}
 \affiliation{Dunlap Institute for Astronomy and Astrophysics, University of Toronto}
\author{Robin Swanson}
 \affiliation{Dunlap Institute for Astronomy and Astrophysics, University of Toronto}
 \affiliation{Department of Computer Science, University of Toronto}
\author{Jacob Taylor}
 \affiliation{David A. Dunlap Department of Astronomy \& Astrophysics, University of Toronto}
 \affiliation{Dunlap Institute for Astronomy and Astrophysics, University of Toronto}
\author{Dan Vargas}
 \affiliation{Steward Observatory, University of Arizona, 933 N. Cherry Ave., Tucson AZ 85721, USA}
\author{Grant West}
 \affiliation{Steward Observatory, University of Arizona, 933 N. Cherry Ave., Tucson AZ 85721, USA}
\author{Jennifer Patience}
 \affiliation{School of Earth and Space Exploration, Arizona State University, Tempe, AZ 85281, USA}
\author{Katie Morzinski}
 \affiliation{Steward Observatory, University of Arizona, 933 N. Cherry Ave., Tucson AZ 85721, USA}

\footnote{Further author information: \\J.A.J.:  jajohnson@arizona.edu, 1 520 621-2288 \\ A.V.:  
 amali@arizona.edu, 1 520 621-2288}

\begin{abstract}
The MMT Adaptive Optics exoPlanet Characterization System (MAPS) is a comprehensive update to the first generation MMT adaptive optics system (MMTAO), designed to produce a facility class suite of instruments whose purpose is to image nearby exoplanets. The system's adaptive secondary mirror (ASM), although comprised in part of legacy components from the MMTAO ASM, represents a major leap forward in engineering, structure and function. The subject of this paper is the design, operation, achievements and technical issues of the MAPS adaptive secondary mirror. We discuss laboratory preparation for on-sky engineering runs, the results of those runs and the issues we discovered, what we learned about those issues in a follow-up period of laboratory work, and the steps we are taking to mitigate them.
\end{abstract}


\section{INTRODUCTION}

The original MMT adaptive optics system, with its 336 voice coil actuator adaptive secondary mirror, was the first of its kind when it was deployed in 2002 \citep{Hinz18}. The original ASM was a cooperative venture between the University of Arizona and the Arcetri Observatory in Italy. Its groundbreaking deformable mirror technology directly led to the development of second-generation adaptive secondaries at the Large Binocular Telescope, the Magellan Telescope, and the future Giant Magellan Telescope.

By the time the MMT system was decommissioned in 2017, it was fully at the end of its functional life. The necessity of retiring the system, as opposed to further extending its life, was obvious from several perspectives. Replacement parts had been exhausted; the capabilities of electronics had improved considerably; and lessons had been learned that would lead to substantially better performance if the system were to be completely redesigned. What was to become the MMT Adaptive optics exoPlanet characterization System (MAPS) was initially funded by an NSF Mid-Scale Innovations Program in Astronomical Sciences (MSIP) seed grant issued in August of 2018, and the process of designing a next-generation Adaptive Optics system began.

Details of the Adaptive Optics System as a whole have been covered in several previous SPIE proceedings \citep{Hinz18,Vaz20,Morzinski20}. The topic of this paper is specifically the MAPS Adaptive Secondary Mirror, including its design, components, operations, and technical issues. Most of the improvements in AO performance brought about by the redesign of the system relate directly to changes in the design of the ASM. Some of these changes are quite substantial, enough so that, following the first generation of ASM design represented by the first MMT ASM and the second generation of ASM design represented by the Large Binocular Telescope mirror, the MAPS ASM is, by all considerations, a third generation adaptive secondary design.

In conceptualizing the improvements desired in the new system, high-level science goals in support of MAPS mission objectives combined with considerations of practicality to create a set of engineering and performance parameters that would inform the design of the new secondary. These can be summarized as follows~\citep{Hinz18,Vaz20}:

\begin{itemize}[noitemsep]
  \item Reduce the power consumption from a typical ASM level of 1000W - 3000W to a level of 200W - 300W during normal operation. The majority of this reduction occurs because of efficiencies in actuator electronics and improved actuator heat extraction. As designed, a single actuator operating under typical seeing conditions should consume around 0.8 W of power, down from 5.4 W. 
  \item Reduced power consumption leads to reduced heat dissipation, allowing for a passive cooling system, obviating the need for any form of active (liquid) cooling system.
  \item Increase operating bandwidth from below 500 Hz to 1000 Hz, and:
  \item Increase speed of positional calculations to 1kHz.
  \item Improve temporal lag by using optimized PID control algorithms to decrease positional settling time by a factor of 2-3, and:
  \item Implement an open loop feed-forward component to actuator control.
  \item Decrease the overall system update rate from 2.7 ms lag to 1 ms in the new system.
  \item Increase the number of corrected modes from 55 for the legacy system to 220 for the new system.  
\end{itemize}

\begin{figure}
  \plotone{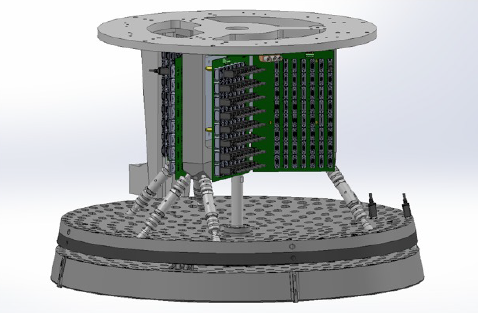}
  \caption{A schematic representation of the MAPS adaptive secondary mirror, showing its primary components, which are identified in the blowup diagram in Fig.~\ref{fig:ASMBlowup} 
  \label{fig:ASMSchematic}}
\end{figure}

With these changes, it was estimated that residual wavefront error could be reduced from between 400 and 500 nms RMS for the old system to 200 nm RMS for the new system, with corresponding gains in Strehl ratio. These changes would also support the more stringent requirements of the MAPS program: higher spectral resolution, increase in achievable contrast, improved image quality over a broader wavelength range, and increased throughput and operational efficiency~\citep{Hinz18}.

\section{DESCRIPTION OF THE MAPS ADAPTIVE SECONDARY MIRROR}

\begin{figure}
  \plotone{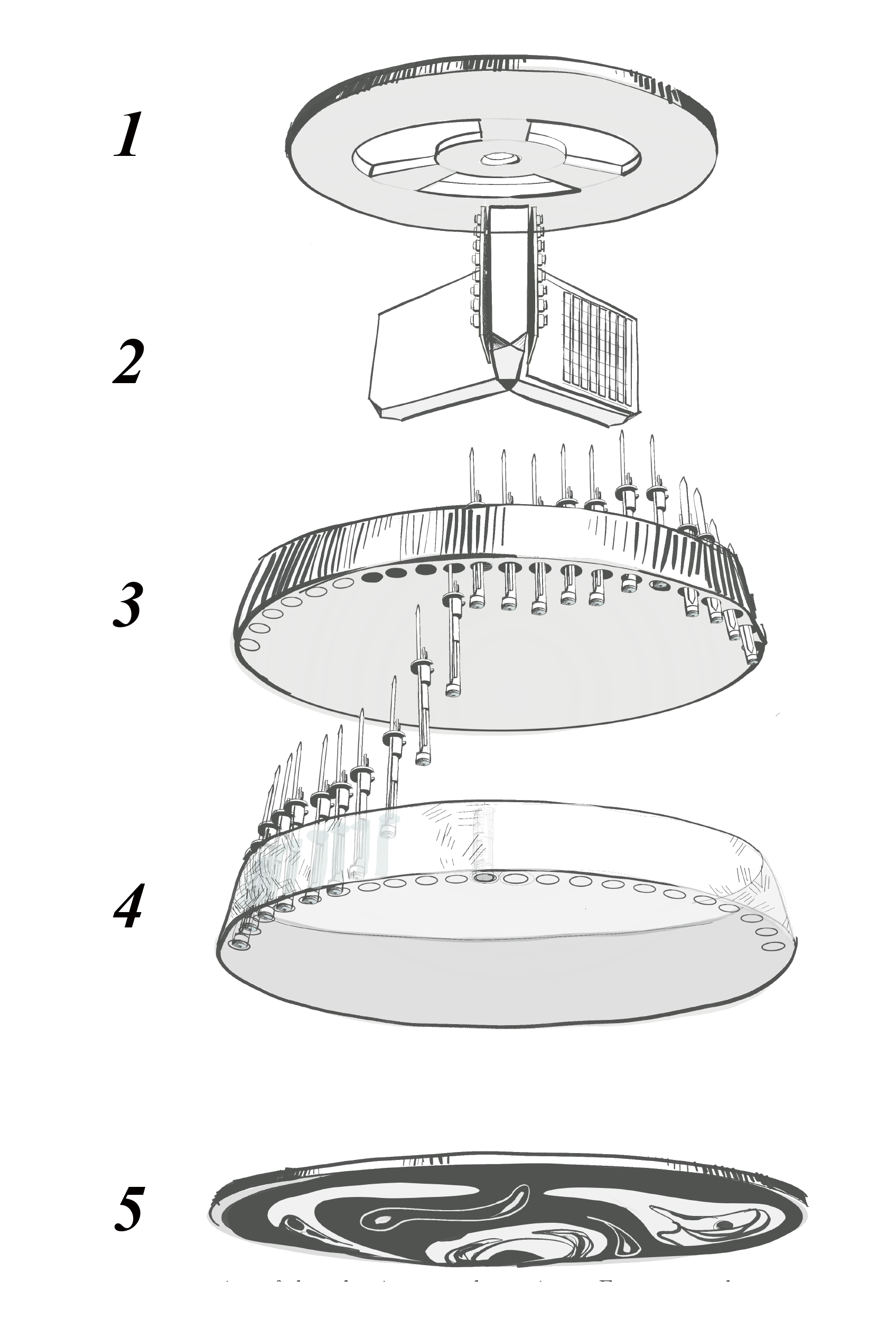}
  \caption{A blowup diagram of the MAPS adaptive secondary mirror, showing its major structural components. 1) is the ring interface, 2) is the central electronics hub, 3) is the cold plate, 4) is the reference body, and 5) is the thin shell.
  \label{fig:ASMBlowup}}
\end{figure}

The MMT telescope is a Cassegrain design with a 6.5-m f/1.25 borosilicate spun-cast primary mirror. The adaptive secondary is an f/15 deformable convex mirror, which provides a 20 arcminute field of view~\citep{West97}. A schematic of the mirror and its support structures is shown in Fig.~\ref{fig:ASMSchematic}, and a blow-up diagram is shown in Fig.~\ref{fig:ASMBlowup}.

\subsection{Structure of the ASM}
 \label{sec:structure}

The following is a review of the mechanical structure and components of the ASM. The MAPS adaptive secondary mirror is essentially an upgrade of the original MMTAO secondary mirror. The cold plate, reference body and thin shell are all legacy components, but the actuators and electronics components are all of new design. Each of these components is discussed below.

\begin{figure}
 \plotone{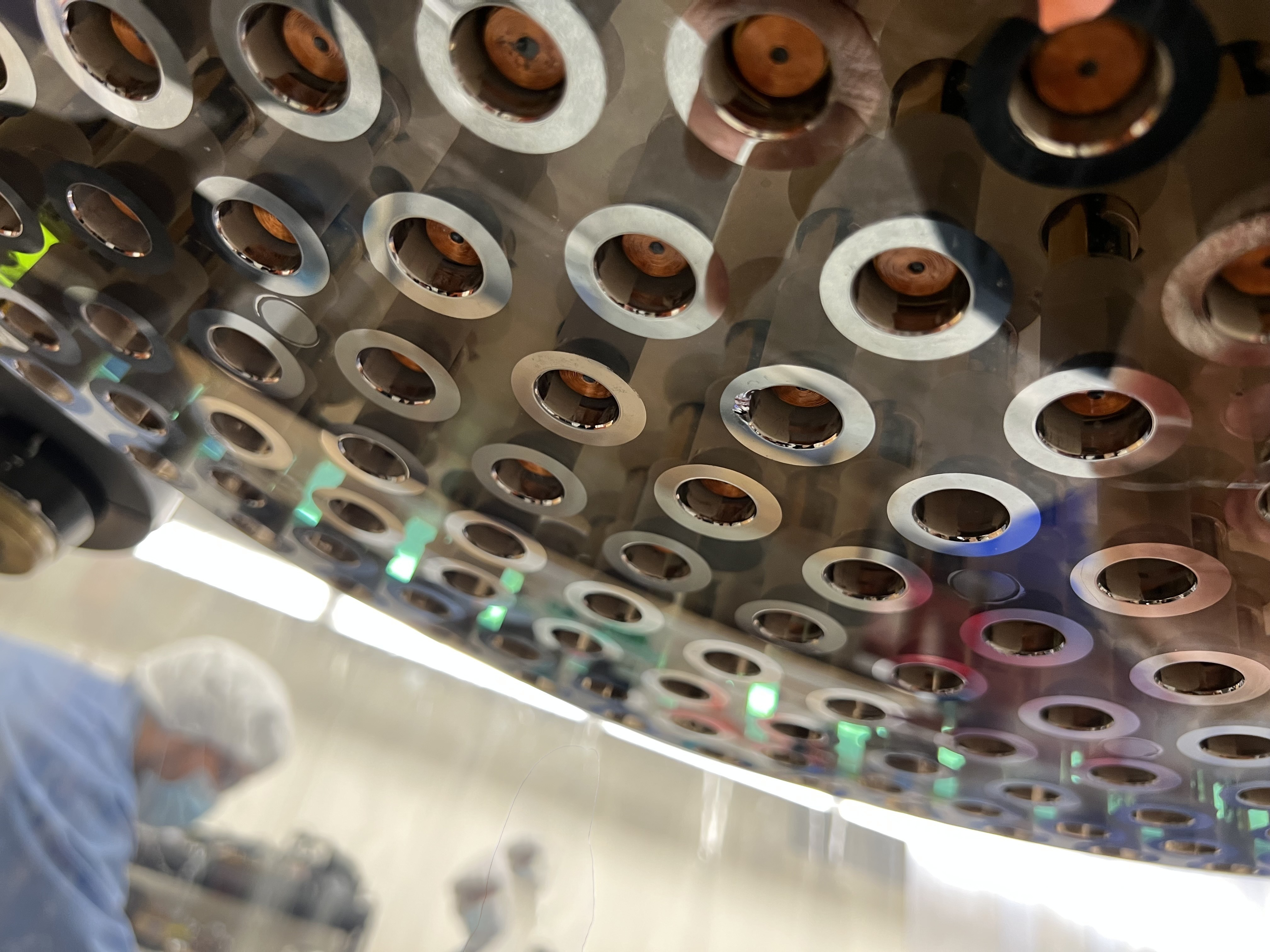}
 \caption{An image of the bottom surface of the Zerodur glass reference body, normally covered by the thin shell, showing the boreholes, actuators and capacitive sensing rings.
 \label{fig:ASMGlass}}
\end{figure}

\textbf{Reference body:} the reference body, so named because it provides a stable reference surface for determining mirror position, is a monolithic 500mm-thick piece of Zerodur glass. It is a legacy component from the MMTAO system that is largely unchanged. 336 boreholes have been drilled through the glass from top surface to bottom surface, and each houses a single actuator.  

The holes are arranged in ten concentric circles, each drilled such that their central axes are normal to the bottom (primary-facing) surface; on that surface, a chromium annulus surrounds each borehole. The annulus (or ring) is an essential part of the capacitive position sensing system; for details of the capacitive sensing system, see Sec.~\ref{sec:capsensing}.  An image of the bottom of the reference body, showing the boreholes with actuators in place and the capacitive sensing rings, is shown in Fig.~\ref{fig:ASMGlass}.

\textbf{Cold plate:} the legacy cold plate is made of a single piece of aluminum and serves as a surface to which the actuators attach. It is no longer used for active cooling but serves as a passive thermal sink. The thermal system is discussed in more detail in Sec.~\ref{sec:thermaldesign}.

\textbf{Hexapod and telescope interface:} a hexapod is attached to the reference body, extends through the cold plate, and is attached at its other end to a metal electronics frame and ring interface. The ring interface is one of the primary attachment points to the telescope. 

\textbf{Electronics hub:} the electronics hub holds the system's housekeeping board, control motherboard, and six daughterboards. Each daughterboard is attached to 56 actuators via USB-C cables. Commands, actuator status info, and monitoring flow between this architecture and a external rack-mounted control computer via Ethernet. 

\begin{figure}
 \plotone{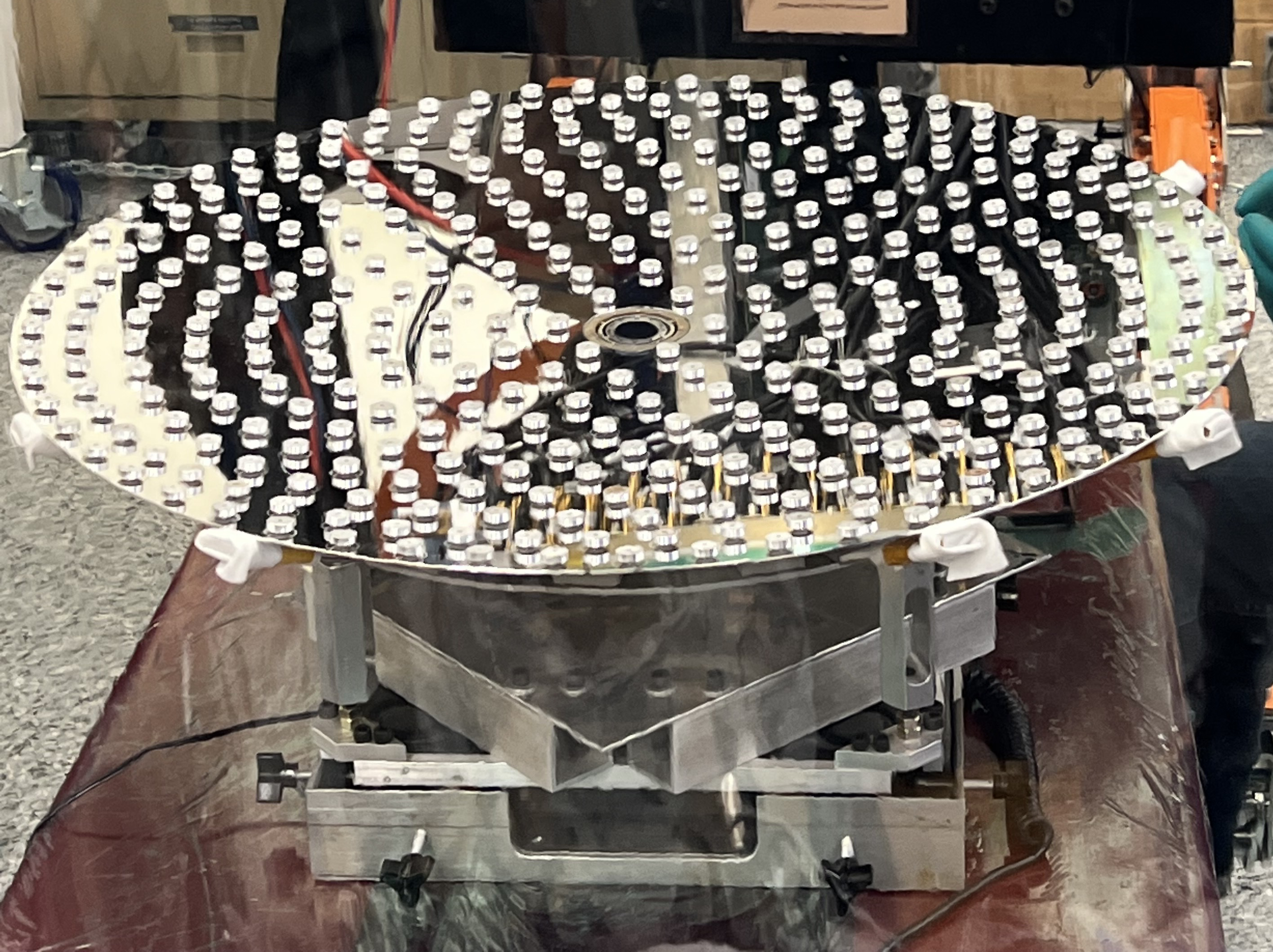}
 \caption{An Image of the top surface of the thin shell (opposite to the mirrored surface), showing the 336 rare-earth magnets that match position to the boreholes in the reference body.
 \label{fig:MirrorMagnets}}
\end{figure}

The motherboard, housekeeping board and daughterboards represent the entirety of the ASM's electronics infrastructure, and this is one of the major improvements that differentiates MAPS from its predecessor: all of the functionality of the electronics that used to be situated outside of the MMTAO ASM have been relocated to the components inside each actuator, allowing, essentially, all of the hardware required for the ASM's operation to be located within the bounds of the ASM itself. 

This is key to the ASM's passive cooling, as it enables each actuator to consume less power and therefore dissipate significantly less heat. For more on the system's power consumption, see Sec.~\ref{sec:powerconsume}.

\textbf{Thin shell:} the thin shell is the secondary mirror's surface. It is 64 cm in diameter, made of Zerodur glass that ranges from 1.8 - 1.9 mm in thickness and weighs 2.6 kg. 

Attached to the upper surface of the thin shell are small, radially polarized rare-earth magnets positioned directly below each of the boreholes; see Fig.~\ref{fig:MirrorMagnets}. The magnetic field generated by the actuator coil interacts with these magnets to provide the force that moves the mirror.

Built into the actuator under the coil is a small bias magnet. These magnets, not present in the original MMTAO, act as a safety mechanism for the fragile thin shell. The attraction between the magnets on the thin shell and the actuator's bias magnet restrains the thin shell from separating from the reference body in the absence of power to the ASM. (The MMTAO design relied on edge clamps and a central retaining ring to restrain the shell). 

\subsection{The MAPS actuators}

Of all the changes to the MMTAO design, the most thorough and profound were those made to the actuators. 

In both the legacy and current systems, the basic procedure is the same: the distance between an actuator and the portion of mirror directly over it is measured by the capacitive sensing system, described in Sec.~\ref{sec:capsensing} below; the difference between the mirror's measured and desired position is calculated and converted to a current value; the current is sent through the coil, creating the magnetic force which in turn moves the mirror.
	
There were three primary design goals for the new actuator design: one, to improve the actuator's dynamic performance; two, to decrease overall power consumption and heat generation; and three, to allow for a configurable control architecture. This was accomplished by moving all of the electronics components required for the actuator's function into the actuator itself. 

The original MMTAO actuator electronics did only two things: take the voltage measurement from the capacitive system, and apply the force to the mirror. The new actuator design consolidates almost every fundamental ASM function to its onboard electronics~\citep{Downey19}. The actuator functions performed at actuator level on MAPS are: 

\begin{itemize}[noitemsep]
  \item Measure the capcitive decay voltage;
  \item Digitize the measurement;
  \item Calculate the force needed to move the mirror to position;
  \item Send the required current through the coil;
  \item Carry out safety and status checks on the actuator;
  \item Synchronize all of the above with the other actuators
\end{itemize}

In the legacy system, conversion of the capacitor's decay voltage measurement to distance, calculation of the required current to move the mirror, monitoring of actuator function, and coordination with other actuators were all accomplished via off-board electronics; now they are built into the actuator itself.

Centralizing the computational function also allowed the use of standardized PCs, instead of customized control circuitry and processors, for external calculations.

\begin{figure}
  \plotone{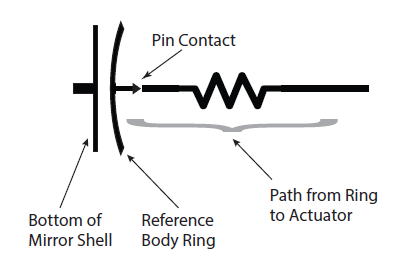}
  \caption{Representation of the capacitive sensing system.
  \label{fig:CapSense}}
\end{figure}  

\subsection{The capacitive sensing system}
 \label{sec:capsensing}

The MAPS ASM calculates the position of the mirror directly over each actuator by utilizing capacitive sensing. The capacitive sensing system uses the fact that a parallel plate capacitor has a known relationship between its geometry and its physical properties, in particular that the voltage between its plates is directly proportional to the distance between its plates and indirectly proportional to its capacitance. This allows us to use the bottom side of the thin shell as one plate, and the circular ring on the reference body as the second plate. As the mirror moves, the distance between plates changes, changing the capacitance.

The capacitive sensing circuit for each actuator is illustrated in Fig~\ref{fig:CapSense}. To read distance, we pulse a voltage (the 'Go' pulse) across the capacitor, which then drops off in a predictable way in the time decay fashion of an RC circuit. We read the voltage at two set times on the drop off curve, subtract the two values, and then convert that to a distance value. In this way, we can read the distance from the reference body to the mirror for every actuator every time the ‘Go’ pulse fires.

\subsection{Power and heat}
 \label{sec:power}

The key to reducing the amount of heat generated by the ASM that must be managed is to reduce the power consumed by each actuator and its supporting electronics. The following sections discuss how MAPS accomplished this.

\subsubsection{Power consumption}
 \label{sec:powerconsume}

By moving functionality from hub and connected electronics to miniaturized onboard components, the total power consumption drops considerably. Table \ref{table:1} shows the comparison between the legacy and MAPS system's power consumption. Although the power consumed by each individual actuator is ten times greater than in the legacy system, the hub electronics consume thirty times less. The substantial reduction in hub power draw decreases hub heat discharge enough to dispense with the necessity of actively cooling it.

\begin{figure}
  \plotone{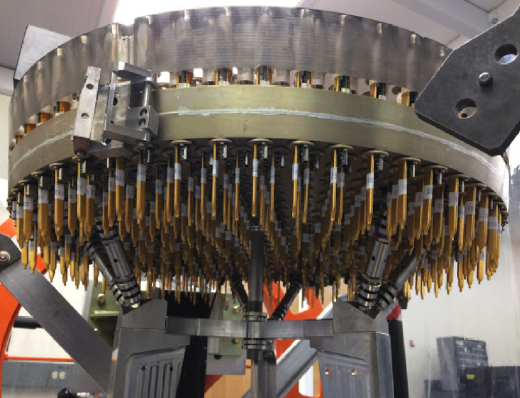}[!t]
  \caption{A view of the top side of the ASM, showing the actuator heat pipes. Note that the actuators do not have their USB cables attached, making it easier to see the whole assembly.
  \label{fig:ThermalPipes}}
\end{figure}

\subsubsection{Thermal system}
 \label{sec:thermaldesign}

The power consumption of the MMTAO legacy system, shown in Table \ref{table:1}, generated a considerable amount of heat and required an active liquid cooling system to dissipate that heat. However, liquid-cooled ASMs are complex and susceptible to coolant leaks (for example, the Large Binocular Telescope had an adaptive secondary coolant leak in 2013). By reducing the overall power consumption, as described above, we reduced the thermal dissipation to the point where the system could be passively air cooled. In this way, The MAPS ASM is the air-cooled Volkswagon of adaptive secondaries. 

The MAPS passive cooling thermal system consists of two components: the cold plate to which the actuators attach (Sec.~\ref{sec:structure}), and a copper heat pipe which is embedded into each actuator. 

A heat pipe has three components: a metal envelope; a working fluid in contact with the envelope which absorbs heat, vaporizes, and moves to the top of the envelope; and a capillary wick which carries the fluid back down the pipe after it release its heat and condenses. They are relatively simple devices but extremely efficient thermal conductors. MAPS heat pipes are copper, 150mm long and flattened, and have an effective thermal conductivity of $\approx30,000~Wm^{-1}K^{-1}$. An image of the ASM, showing the actuator thermal pipes, is shown in Fig.~\ref{fig:ThermalPipes}.
   
What little heat is generated by the new hub architecture of motherboard, daughterboards, and housekeeping board is transferred to the surrounding ambient air. 

\section{INITIAL LABORATORY TESTING AND CALIBRATION}
 \label{sec:InitialLab}

Beginning in December 2019, we assembled the ASM from parts, verified its operation, and performed the first rounds of calibrations in a laboratory environment. As discussed in detail in \cite{Vaz20}, we were successfully able to demonstrate basic ASM functions: to float the shell at a mean gap of $35\mu m$; to apply a set of new positions to all actuators; and to optically confirm the result. 

\begin{figure*}
  \plotone{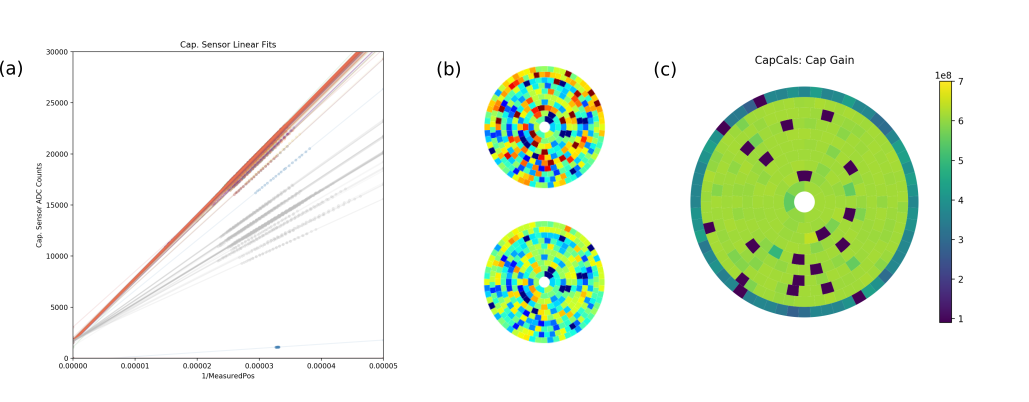}
  \caption{Demonstration of capacitive sensor calibrations. (a) The ``refinement" step, during which linear fits between ADC counts and 1/distance are used to better estimate cap sensor parameters. (b) Before (top) and after (bottom) that refinement, applied to the same smooth mirror shape. Red actuator high spots turn yellow or green as better calibrations allow the reported position to closer approach the actual physical gap size. (c) Variation of one of the capacitive sensor parameters across the shell. Because the calibration depends on the area of the chrome armature, and because the outermost ring of actuators has a clipped armature, the outer ring calibrations are significantly different from those within.
  \label{fig:CapCal}}
  \vspace{5mm}
\end{figure*}

We made rudimentary capacitive sensor calibrations -- mapping the sensor's raw ADC counts to physical distance -- using the geometry of the double-plate capacitor. Those we refined further with linear fits to sweeps across multiple gaps (see Fig~\ref{fig:CapCal}). To tie our numbers to the physical world, we checked the gap at the edge of the shell with a plastic shim. That absolute gap measurement is imprecise by perhaps as much as +/-5 $\mu$m, because we had limited shim sizes. However, because it is the \textit{relative} positions of the actuators that determine the shape of the shell, that margin is acceptable. 

We were also able to perform rough tuning of the actuator loop and to generated a very rough feed-forward matrix. We will need to refine data acquisition and analysis for both of those before the results are ready for usage on sky, but the preliminary versions were enough to show that the underlying techniques are sound.

\begin{table}
    \hspace{-1cm}
    \begin{tabular}{|l|l|l|} 
      \hline
        \textbf{Source} & \textbf{Legacy} & \textbf{MAPS}  \\
      \hline
        Ind Actuator Electronics & 0.03 W & 0.33 W   \\
      \hline
        Individual Coil Power & 0.33 W & 0.41 W   \\
      \hline
        Individual Total & 0.36 W & 0.74 W   \\
      \hline
        Total Across All & 120W & 249 W  \\
      \hline
        Hub Electronics & 1680 W & 50 W  \\
      \hline
        \textbf{TOTAL} & 1800 W & 299 W  \\    
      \hline 
    \end{tabular}
    \vspace{4mm}
  \caption{Comparison of the Legacy and MAPS Systems \\ Power Consumption}
  \label{table:1}
\end{table}

\subsection{Preparing for first light}

More recently, in preparation for full system integration and on-sky science, we have been taking the very first steps beyond basic operation (``make it work") towards refining functionality (``make it work well"). For eventual full MAPS AO performance, the ASM must meet the following specifications: 

\begin{enumerate}[noitemsep]
  \item \textbf{Shape}: the mirror must be able to deform into at least 160 modes. The finite number of corrected modes is a strict limit on wavefront correction: we cannot correct a shape we cannot make.
  \item \textbf{Settling time}: Each actuator will settle into position within 1ms after it is commanded to move.
  \item \textbf{Measurement noise}: Each actuator shall have measured position RMS no greater than 10nm.
\end{enumerate}

On sky, actuator noise translates directly into increased wavefront error; off sky, it impedes much of the critical calibration work, yielding suboptimal actuator performance. We discuss actuator positional noise further in Sec.~\ref{sec:noise}.

\subsubsection{Position commands and modal control}
 \label{sec:Modes}

\begin{figure*}
  \plotone{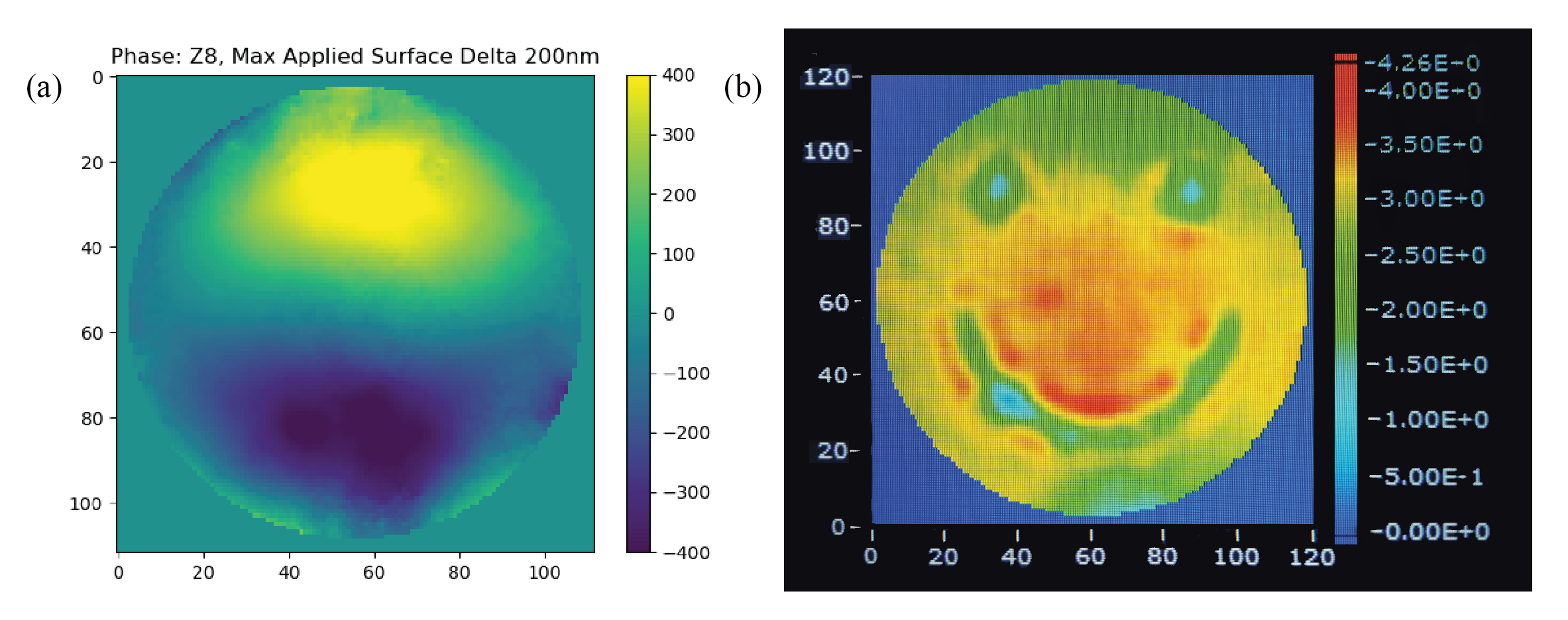}
  \caption{Demonstration of successful control over the ASM surface as viewed by a PHASICS camera. (a) Modal control, here displaying Zernike mode 8 (coma). (b) Control over an arbitrary set of single actuators, here showing the human propensity for anthropomorphism.
  \label{fig:SurfaceControl}}
  \vspace{6mm}
\end{figure*}

All commands are sent to the ASM as a 336-element vector of absolute positions. The overall shape, though, is first constructed from a linear combination of modes from a suitable basis set. To verify that the ASM can make the shapes we need it to, we applied commands on both the single-actuator and the modal-basis levels and viewed the resulting wavefront optically, from our test bench. Illustrative results are shown in Figure~\ref{fig:SurfaceControl}.

Shapes built from modes, although necessarily approximate, become better approximations as the number of modes increases. In principle, a 336-actuator mirror can produce 336 orthogonal modes. In reality, the combination of actuator failures and measurement noise means that 300 modes is a more feasible ``goal maximum", and 160-220 is enough for typical science in the MAPS program. 

In our experience, there is a trade off between precision correction and system stability: the highest number of modes can produce the most precise correction, but high spatial frequency shapes are more likely to trip the ASM safety limits. At telescopes with similar AO systems (LBTI, MagAO), we have had good success tailoring the correction ``flavor" to the science requirements of the program at hand as well as the environmental conditions of that particular night, and that allowing operator choice between multiple setups yields more science output. We expect that the same will hold true for MAPS.

For testing and early on-sky work, we have been using Zernike modes as our basis for building mirror shapes: they serve the purpose fairly well, are easy to visually identify for sanity checks, and are straightforward to implement. However, a modal basis more tailored to the physical properties of our system -- in particular, the natural resonant modes of the thin shell -- would allow us to better approximate the desired shape without increasing the number of modes. Mirror modes are determined as part of the feed-forward matrix generation procedure, and we do plan to implement them later. We can expect a substantial gain in performance when we do so.

\subsubsection{Settling time}

Processes that seem instantaneous on a human scale become significant on the scale of a single millisecond. A command sent from the control computer, for instance, takes some small but finite time to traverse the electronics on its way to the ASM; the actuators take some time to process it as a new setpoint; the PID loop takes some time to move an actuator and its associated block of shell; the shell itself may need additional time to damp any vibration induced by that movement. 

What we are really interested in is the time for the entire shell to achieve a new shape, not just a single actuator, but because of the way our telemetry is set up, we can only sample \textit{either} a single actuator at high speed (once every 30$\mu$s) \textit{or} all actuators at the default rate (once every 900$\mu$s). So, for measurement purposes, we define the single-actuator ``settling time" as the time elapsed between (a) when an actuator gets a new position command, and (b) when it enters and remains within 10$\%$ of that command.

With solely the actuator PID loop governing their motion, only a handful of actuators on the very outermost ring of the unit can achieve the 1ms mark. Those in the other nine rings, matched to much stiffer regions of the shell, cannot easily achieve the high spatial frequency motion that a single-actuator ``poke" represents. 

The standard solution, with the previous incarnation of MMTAO and with similar ASMs since, has been the introduction, in parallel with the PID feedback loop, of a feed-forward matrix: a two-dimensional lookup table that allows us to calculate the approximate coil current required, and send that directly to the actuators, so that they start closer to the desired position.

\begin{figure*}
  \centering
  \includegraphics[height=15cm]{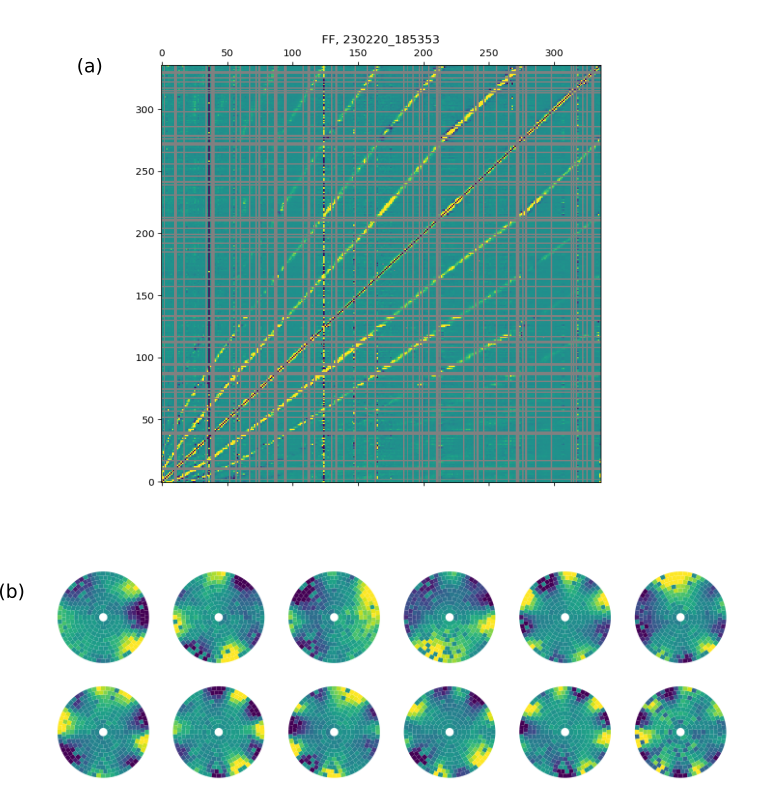} 
  \caption{An illustration of the negative impact of actuator noise on feed-forward matrix calculations. (s) A feed-forward matrix. Actuators that are either non-functional or known to be especially noisy have been greyed out, but even so, several more noisy columns remain. (b) The effect of noise on modal decomposition. With so much noise, even these first 12 decomposed modes show unevenness and propagated noise in their patterns.
  \label{fig:FFNoise}}
  \vspace{5mm}
\end{figure*}

Figure~\ref{fig:FFNoise} shows our early attempts at feed-forward matrix generation. We achieved some limited success with single-actuator response times, but full use of the feed-forward functionality has been hampered, again, by high and unpredictable actuator noise. 

Further, because the feed-forward sends coil current directly to the actuators, an unwary operator or mistyped command could easily dislodge the shell from the reference body entirely, or even launch it with sufficient speed to break the last-resort restraining clips and shower its shattered self upon the telescope primary. We do not yet have software checks in place at that level, although they are planned. Further discussion of software safeguards is in Section~\ref{sec:softwaresafety}.

We have deferred work on the feed-forward matrix until non-feed-forward operation has been thoroughly tested on sky, and safety systems are in place.

\subsubsection{Current on-sky operational state}

For these first on-sky runs, we have limited the unit to a subset of its full functionality. In particular:

\begin{itemize}[noitemsep]
  \item we generally use an overall loop speed of 500Hz, rather than 1kHz;
  \item we operate entirely without the feed-forward matrix;
  \item we use fairly conservative values for the actuator loop gains;
  \item we use a mean gap of only 35 or 40$\mu$m, both of which are small enough to take advantage of air damping as a built in safeguard against excess oscillation
\end{itemize}

Together, those settings allow us enough functionality to check out integration with the wavefront sensors and telescope, but we do trade wavefront correction performance for safety and stability of the ASM and observatory. 

As we gain better understanding of how the ASM performs within the full AO system and in the telescope environment, we plan to implement the rest of the features that allow for optimal AO correction: full 1kHz loop speed, feed-forward implementation, tighter actuator tuning, and possibly a selection of larger gap configurations. 

\subsection{Actuator failure}

Actuators can ``die" in many ways, temporarily or forever, with fixes that may be easy, difficult, impossible, or unknown. 

An early and persistent cause for concern has been our apparent actuator attrition rate. Several of the most common actuator deaths are described in Table~\ref{tab:ActuatorFailures}.  We discuss actuator failure modes and their possible causes further in Sec.~\ref{sec:taxonomy}.

\begin{table*}[t]
  \caption{Common failure modes of individual actuators.} 
  \vspace{3mm}
  \hspace{-4mm}
  \label{tab:ActuatorFailures}
  \begin{tabular}{|p{2.1in}|p{2.1in}|p{2.1in}|} 
    \hline
    \rule[-1ex]{0pt}{3.5ex}   \textbf{Observed Failure} & \textbf{Description/Cause}  & \textbf{Fix}  \\
    \hline
    \rule[-1ex]{0pt}{3.5ex}   Reports 0xDEADBEEF & Actuator failed its checksum test; low-level board problem & Replace actuator \\ 
    \hline 
    \rule[-1ex]{0pt}{3.5ex}   Reports 0xDEADFEED & Broken communication between command computer and actuator & Check cables or replace actuator \\ 
    \hline
    \rule[-1ex]{0pt}{3.5ex}   Measured position at max value & Cap sensor problem & Replace Actuator \\
    \hline
    \rule[-1ex]{0pt}{3.5ex}   Persistent WildCoil errors & Genuine safety issue, Overtuning Config problem, Bad commands & Check parameters, lower PID gains, re-calibrate \\
    \hline
    \rule[-1ex]{0pt}{3.5ex}   Newly noisy measurements & Unknown & Unknown \\
    \hline
   \end{tabular}
 \label{table:ActFailures}
 \vspace{2mm}
\end{table*}

\subsection{Optical testing}

The secondary mirror of a Cassegrain telescope does not have an intermediate real focus. Consequently, we cannot make use of techniques developed at, for instance, LBTO, where an artificial star placed at one focus allows for off-sky testing and calibration. Instead, where possible, we use an optical test stand designed for our unit, and a PHASICS SID4 unit to image the wavefront after it has twice reflected off the ASM.
   
That system has some limitations: 

\begin{itemize}[noitemsep]
  \item Rough alignment involves manual fine movements of the 600kg ASM unit, which can be difficult; 
  \item We have no easy way of introducing an outside reference flat; \item We need to balance airflow for actuator cooling against optical requirements for a vibration-quiet environment;
  \item Neither the ASM nor the test stand optics can, by themselves, generate an unreferenced optical flat. 
\end{itemize}

Many incremental improvements to the test stand, detailed in \cite{Montoya22}, have largely countered the first three issues, but the lack of a good reference flat plagued us all the way through to sky time. As ``flattish" shapes, we tried a handful of different strategies:

\begin{itemize}[noitemsep]
  \item the \textbf{electronic-position flat}, made by commanding all actuators to the same electronically-measured position, and smoothing currents there. Not necessarily physically flat -- susceptible to bad cap sensor calibration and to any error of reference body curvature; requires high coil currents to maintain an exact shape.
  \item the \textbf{force flat}, made by trying to minimize deviations in coil current across the shell. Has low power requirements and consequently low heat generation, but how close is the natural curvature of the shell to an optical flat for the installed unit?
  \item the \textbf{piston flat}, made by starting with the shell held against the ref body and gradually pistoning it to a reasonable gap position. Unfortunately, preserves the ``island" signatures of tiny dust particles that are otherwise small enough not to interfere with operation.
  \item the \textbf{pretty-looking flat} made by an operator sending single actuator commands by hand, trying to make the raw PHASICS image as smooth as possible. Neither efficient nor effective.
\end{itemize}

In the end, once on-sky, none of the above were used except as the most rudimentary of starting points.

\begin{figure}[b]
  \plotone{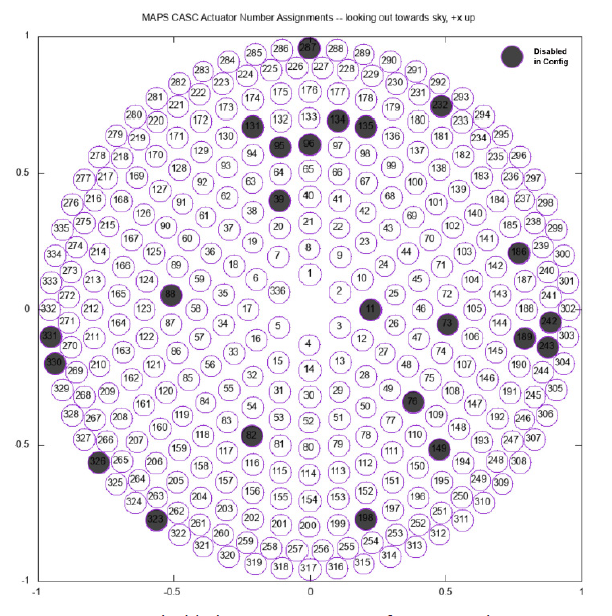}
  \caption{Actuator map at beginning of first engineering run, showing actuators disabled in the configuration files.
  \label{fig:DeadActuators01}}
\end{figure}

\section{TIMELINE OF MAPS ON THE SKY}

\begin{figure*}[t!]
  \plotone{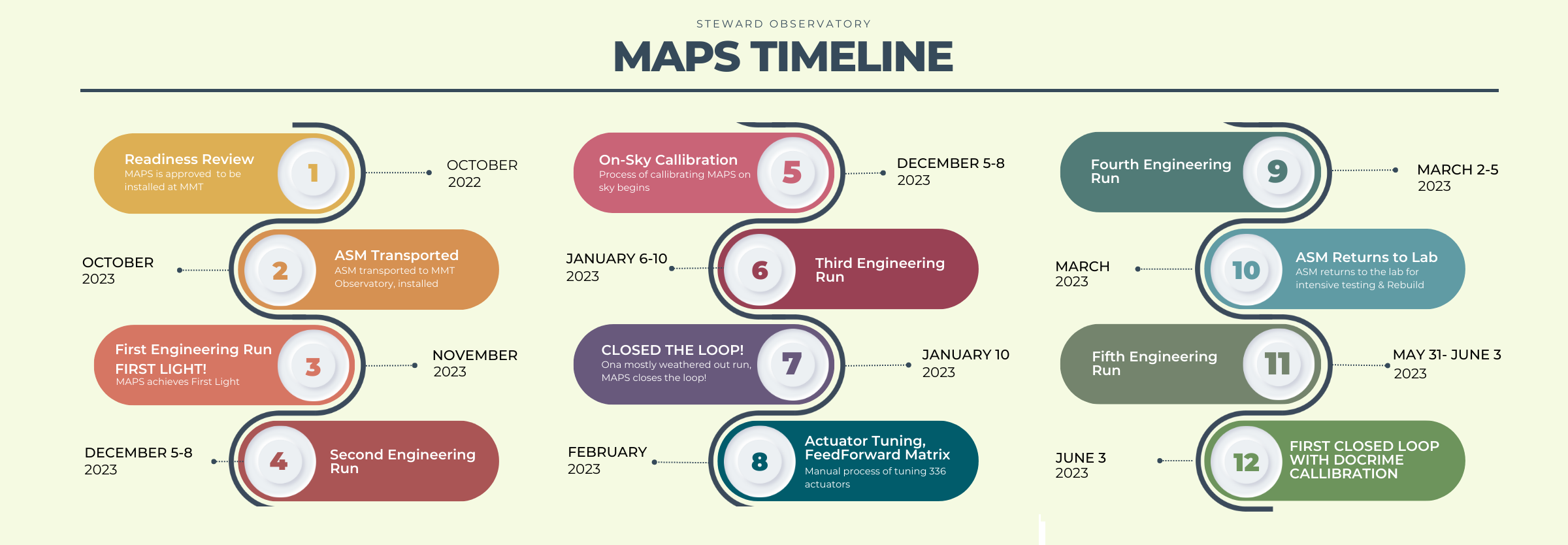}
  \caption{A timeline of major events in the progress of MAPS since the readiness review. The numbers on the timeline are keyed to the subsections below.
  \label{fig:timeline}}
  \vspace{4mm}
\end{figure*}

Due to the issues discussed above in Sec.~\ref{sec:InitialLab}, attempts at optically flattening the mirror had not succeeded; the ASM debuted with a ``force-flat". Furthermore, 23 actuators had already been disabled in the configuration files due to various errors detected in the lab. A map of the actuators showing those that had been disabled is shown in Fig.~\ref{fig:DeadActuators01}.

The problem with taking actuators out of service is that they are not neutral floaters: because they have bias magnets, they effectively become a drag on the actuators immediately around them. Those actuators must then struggle to reach their commanded positions. This, in turn, requires those actuators to produce more force, and therefore draw more current. We call this the \textit{Proximity Effect}, and finding ways to mitigate this effect is a priority in future work. We discuss the effect of these actuators in section \ref{sec:spectra}.

The subsection numbers in this section are keyed to the timeline numbers in Fig.~\ref{fig:timeline}. 
   
\subsection{Readiness review} 

The MAPS project passed its readiness review in October 2022, and was approved to begin a series of four engineering runs at the MMT telescope. All aspects of the MAPS system were examined and discussed, with mirror safety being of primary importance. It was agreed that MAPS should enter its next phase.

\subsection{ASM transported}

By the 27th of October, the ASM had been transported successfully to the summit of Mt. Hopkins and mounted on the telescope. By October 31st, the rest of the MAPS system, including the top ring, the PISCES camera, and all associated servers and cabling had been completed. 

\subsection{First engineering run, first light}

\begin{figure}[!b]
  \plotone{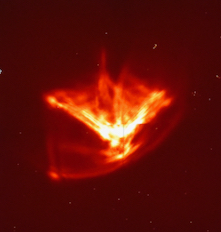}
  \caption{MAPS first light image of Beta Tauri. Image taken in $K$-band with the PISCES wide field infrared camera. This image is the result of the uncorrected shape of the ASM, without closed-loop AO control.
  \label{fig:firstlight}}
\end{figure}

On November 1, MAPS saw first light on Beta Tauri, as shown in Fig.~\ref{fig:firstlight}. As expected, the image is out of focus and heavy in aberration, partially a result of the mirror not being optically flattened. This issue was to be problematic over the next several runs. Calibration of a convex secondary on a Cassegrain configuration is difficult as it is; without a reasonably flattened mirror, it cannot be done. So we immediately began to try to flatten the mirror.

One of the first ways we attempted to do this was by using a poke and optimize routine, i.e.,  poke individual actuators, see the positional response, minimize distance, iterate. The idea is to get as flat a figure as possible, and then eventually use the wavefront sensors to do the rest. 

The mirror was not having it, however (oh, no you don't!). Almost as soon as we started sending positional commands, the mirror balked by drawing large amounts of current and frequently safing itself (\textit{'safing'} is when the mirror puts itself into a standby mode so as to avoid damaging itself). 

It was behaving as if the amplitude of the poke commands were excessive, but this was not the case. The poke routine was sending 10-20 nm amplitudes. Even at this level, the mirror exceeded its specified power consumption by a factor of two. (This was eventually found to be a units issue with competing software packages.)

\subsubsection{A note on actuators behaving badly: wildcoil}

Another unexpected discovery was that actuators were frequently going 'wildcoil'. A \textit{wildcoil} event is triggered at the actuator level, when an actuator draws a large amount of current in an oscillatory manner... positive current, then negative current, repeat... within a specified time period. 

The intent of the wildcoil mechanism is to protect the mirror against large runaway vibrations at one of its resonant frequencies, which would pose a genuine danger to the integrity of the shell. In other words, wildcoil is the final safety check keeping a thin shell from becoming shattered glass. 

An ASM run within its limits should NOT produce regular wildcoil alerts. The simple existence of a wildcoil does not mean the actuator has failed, or is failing, but it does indicate that something about its configuration or its use is not ideal. Many such alerts in rapid succession should be a red flag. We took them as such.

In one evening over 30 actuators triggered wildcoil alerts. This requires operations to be halted, the ASM powered down, and the actuator to be disabled. These are not permanent deactivations, however. Upon reactivation, an actuator will usually resume normal function.

\subsubsection{More actuators bid adieu}
These were not the only issues we encountered. By the end of the first run, another twelve actuators had to be disabled for various reasons. This was a remarkable number given that three of the four nights were almost completely lost to cloud cover. These errors, such as the inability of the actuator to sense its own position, becoming electrically non-responsive, and overheating, are not errors actuators recover from. When disabled for this reason, they must be replaced, which requires dis-assembling the mirror in the lab. And as more actuators become disabled, the proximal effect magnifies, and the mirror draws more current. 

\subsection{Second engineering run}

The second run, December 5-8, 2023, started with 38 disabled actuators and only marginally better weather. We achieved first pupils on the IR Wavefront sensor, shown below in Fig.~\ref{fig:pupils}.The mirror continued to behave similarly to the first run, with six more actuators disabled due to fatal errors, large power draws, and an increasing number of wildcoil events (over 100). Even given this, progress was made in improving flats. This progress allowed us to make attempts at building an interaction matrix. 

\subsection{On-sky calibration}

Typically, interaction matrices are generated by using a calibration source at the entrance focus of the AO system, but the Cassegrain configuration has no intermediate focal plane. We attempted to generate interaction matrices in two ways. The first was using the on-sky generation methodology that is inherent to the CACAO software. The second involved generating random patterns on the DM and using the AO telemetry stream to build the interaction matrix. This is the `DOCRIME' method~\citep{Lai20}. Our observation has been that the DOCRIME method is robust (it created a usable interaction matrix using a questionably flattened mirror) and generates results fairly quickly.  By the end of the run, we had generated a functional interaction matrix through CACAO. 

\subsection{Third engineering run}

The January run started with 44 disabled actuators, leaving 292 in service. A large portion of the run was dedicated to attempting to create a usable flat. We started by using gradient descent image sharpening, which gave us a baseline flat. As the run went on, we modified this essentially by eye. We used the real-time image of the pyramid pupils, and started adding Zernike modes to the mirror with the control system's built-in Zernike generator, as discussed in Sec.~\ref{sec:Modes}. The goal was to get as close to evenly illuminated pupils as possible. We adjusted the mode and amplitude through the generator until portions of the pupil filled in, then added that mode to the flat. This is a form of chi-by-eye adjustment, but it served its purpose. 

\begin{figure}[t]
  \plotone{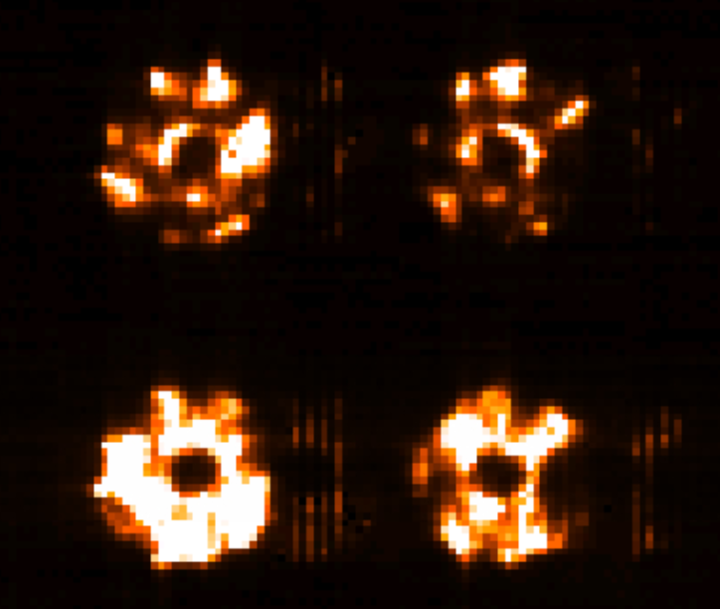}
  \caption{MAPS first IR Wavefront Sensor pupils.
  \label{fig:pupils}}
\end{figure}

\subsection{Closed the loop!}

On the evening of 10 January 2023, with clouds in the sky, the MAPS system closed the loop for the first time. It is a testimony to the design of the ASM and the control software, and the robustness of the wavefront sensor design, that, even though a substantial number of actuators had been disabled, a flat had been created mostly by eye, there were clouds in the sky, and only a rudimentary interaction matrix was in place, MAPS closed the loop. The system ran at 200Hz; the object was a 3.12 magnitude A7 star, and the loop was closed on 64 modes.

\begin{figure*}[t!]
 \plotone{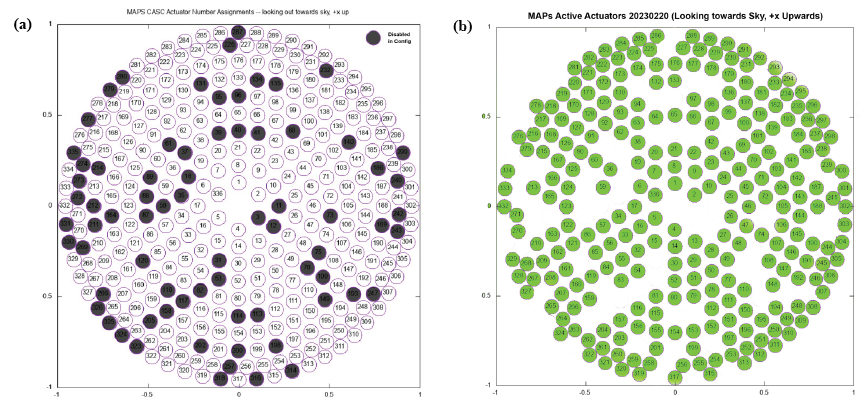}
 \caption{Actuator map of disabled actuators as of the end of the fourth engineering run. (a) shows disabled actuators in black; (b) shows active actuators.
 \label{fig:deadactuators02}}
\end{figure*}

\subsection{Interregnum: actuator tuning}

A further 19 actuators had been disabled during the third run, for a total of 63, leaving 273 of 336 in service. We decided that, in the break between the third and fourth engineering runs, we would attempt to get as much functionality out of those remaining actuators as possible, and attempt to understand the causes of escalating actuator attrition. We decided to proceed on two fronts, each of which we expected to have significant impact on system performance: actuator tuning, and creating a feed forward matrix. In the process, we would examine every remaining actuator in an attempt to identify those that were solidly healthy, and those that were giving signs of impending issues. (Actuator tuning and associated topics are discussed in Sec.~\ref{sec:actcon}.)

We succeeded in tuning all functional actuators and creating a feed-forward matrix. Although we purposely kept our tuning values to moderate levels, we saw immediate gains in system latency during the fourth run. We didn't utilize the feed-forward matrix, however, due to software control concerns.

We also re-enabled all previously disabled actuators, tested them, and examined their power spectra (see section \ref{sec:spectra}). The results of this process, combined with the consolidation of historical manufacturing and initial testing information, led to the creation of the MAPS actuator database. The database contains all the information we have on every actuator in our stock, including those in use and those stored as replacements. For those in use, it contains power spectra, wildcoil counts, history of failure, noise levels, etc. We are currently using the database to quantify indicators of imminent actuator failure.

\subsection{Fourth engineering run}

During the fourth run, 2-5 March 2023, the actuator wildcoil count reached 90 events in 30 minutes. Actuator noise was in some cases in excess of 50 nm RMS (see Section \ref{sec:noise} for a discussion of noise). Issues with integration of the Chai, CACAO and INDI software (see Sec.~\ref{sec:softwaresafety}) were causing unexpected high-amplitude commands to be sent to the mirror, causing either the software or the mirror operator to hit the panic button. The mirror was having difficulty with reaching and holding commanded positions, especially in its outer ring actuators. The actuator disabled count had reached 73 out of 336, 22\% of the total. The actuator map now looked like Fig.~\ref{fig:deadactuators02}.
   
We decided to take the mirror back to the lab, disassemble it, and examine it. At the same time, the software team would work on integrating the various software components and would build safety routines to protect the mirror. We gave ourselves two months to determine why the ASM was having such unexpected issues.

\subsection{ASM returns to lab}

During the second week of March, the ASM was transported from Mount Hopkins to the Steward Observatory, mounted on its test stand, and the thin shell was removed. The reference body was visually inspected and measurements were taken of the actuator depth in the boreholes and the axial alignment of the actuator with the hole. The cabling was removed from each actuator and inspected. 

Many of the actuators that had reported being electrically dead were found to be either simply disconnected due to stress in the USB cabling or had loose or cracked connectors. Others were removed for replacement. We found a correlation between actuator position in the boreholes and actuators that had reported certain types of errors; this is discussed in Sec.~\ref{sec:causes}. Based on this, we made determinations on actuators that had reported `Measured Position' type errors as to whether they should be replaced. Two of the actuators that had reported persistent large current draws and overheating were examined electronically and found to show no immediate indicators of failure. 

The ASM was reassembled during the third week of May. 24 actuators were replaced out of backup stock. The axial and vertical alignment of each actuator was corrected if needed, and steps were taken to reinforce their positions with Teflon spacing rings. Cabling was replaced with longer lengths for all instances where reattaching the cable created stress at the connectors. The thin shell was reinstalled, and the mirror was tested for noise and actuator functionality. No actuators reported fatal errors at start-up, although 12 were disabled for inspection. The mirror was then prepared to return to MMT.

\subsection{Fifth engineering run}

The fifth engineering run, held May 31st to June 3rd, represented a substantial improvement in performance and capability for the adaptive secondary mirror. wildcoil events rarely occurred. Noise levels had been substantially reduced, but not to the desired specification levels. The current draw remained well within expected levels. The safety protections in the software worked as expected. We rebuilt flats using the gradient descent methodology augmented by visual pyramid pupil inspection and Zernike flat building, and we successfully created an interaction matrix via the DOCRIME method.

\subsection{Second closed loop}

On June third, MAPS again closed the loop, at 500 Hz, using the DOCRIME interaction matrix. This ended the first series of MAPS engineering runs on a positive note.

\section{ISSUES AND SOLUTIONS}

The MAPS adaptive secondary mirror is a cutting edge device with a host of new technology under the hood. In this early stage of use, it is closer to a prototype than a science-ready instrument, but that is rapidly changing as we identify problems, diagnose them, and make improvements.The actuator design is inherently sound and functional. Most of the issues we see come from small but correctable design choices. Noise is currently an issue; but we've seen the ASM perform to specification (see Sec.~\Ref{sec:tinhat}. Actuators slipping from their mounted positions in bore holes cause issues in actuator performance, but are correctable. Lag time and system frequency are below expectation, but the system is missing two principle components... tuning and feed-forward. This section discusses the most important of those issues and our attempts at addressing them.

 \begin{figure*} [!ht]
  \plotone{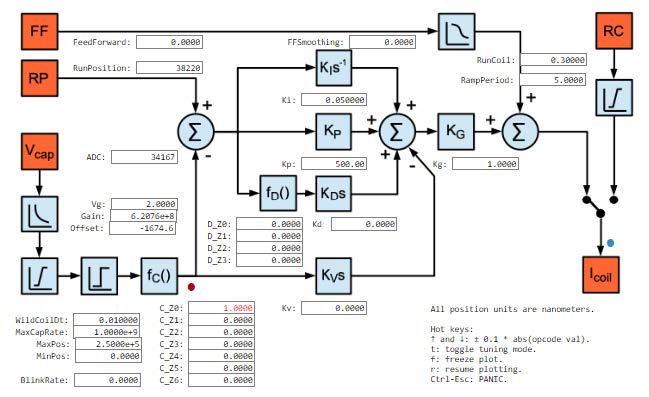}
  \caption{The ‘Actuator Explorer’, or actuator control page, from the MAPs control interface.
  \label{fig:pidcontrol}}
  \vspace{5mm}
\end{figure*}

\subsection{Actuator control}
 \label{sec:actcon}

The actuators of the MAPS ASM are each controlled by their own modified Proportional-Integral-Derivative (PID) control system. The operator has access to each actuator’s control system through the web GUI ‘Actuator Explorer’ page, an example of which is shown in image Fig~\ref{fig:pidcontrol}.

\subsubsection{PID control system}

The MAPS control system is a \textit{modified} PID control. In addition to the standard proportional, integral, and derivative variable, there is also a Velocity Dampening Gain variable and an Output Gain variable. For most of the first five runs, we had been running the mirror essentially untuned, using minimum values for the Proportional and Integral variables, setting the Derivative and Velocity Dampening variable to zero, and setting the Output Gain to one. 

We have hesitated to push too far forward with tuning, because tuning noisy actuators gives incorrect results. As discussed in Sec.~\ref{sec:methodoftuning}, tuning actuators requires adjusting their proportional gain until the actuator begins to oscillate. If the actuator is noisy, its jitter makes it hard to determine whether the actuator is oscillating due to adjusting the variable, or due to noise.

\subsubsection{Feed forward system}

Another essential element of actuator control is the `Feed-Forward' system, consisting of a feed-forward Matrix and a feed-forward bias file. The feedforward system is essentially predictive: it decreases the time it takes for an actuator to reach the commanded position by sending an estimate of the current required, then allowing the PID system to do the final adjustments. 

\subsubsection{Method of tuning}
 \label{sec:methodoftuning}

The process we use to tune actuators is called the Ziegler-Nichols (ZN) PI method.

To start the tuning process, we decided to adjust only the proportional and integral variables, $K_p$ and $K_i$, respectively.  We start with $K_i$ and $K_d$ set to zero and begin increasing $K_p$ until the actuator begins to oscillate. We can tell when the actuator oscillates by looking at its power spectrum. (Power spectra are becoming the tool of choice for understanding actuator behavior and health; see Sec.~\ref{sec:spectra}.)

When the power spectrum indicates that the actuator is oscillating, we stop increasing $K_P$ and note its value. This becomes the 'ultimate gain' $K_u$. Next, we determine the period of oscillation $T_u$. We then take $0.45K_u$ and set this as the value of $K_P$. $K_i$ is then set to $0.54K_uT_u^{-1}$. 

When we begin to tune using all three tuning parameters, the process is the same, except that the final values become $K_p = 0.60K_u$, $K_i = 1.2/K_uT_u^{-1}$, and $K_d = 0.075K_uT_u$.

\subsubsection{Power Spectra}
 \label{sec:spectra}

One very useful diagnostic, in tuning and otherwise, is the plotted power spectrum of an actuator's measured position. The spectrum is the Fourier transform of the actuator's position time series, and tells us how much oscillation this actuator measures, at each given frequency. (Note: the telemetry system collects data at ~1100Hz, so the maximum detectable frequency is 550Hz.) 

\begin{figure*}[t!]
  \plotone{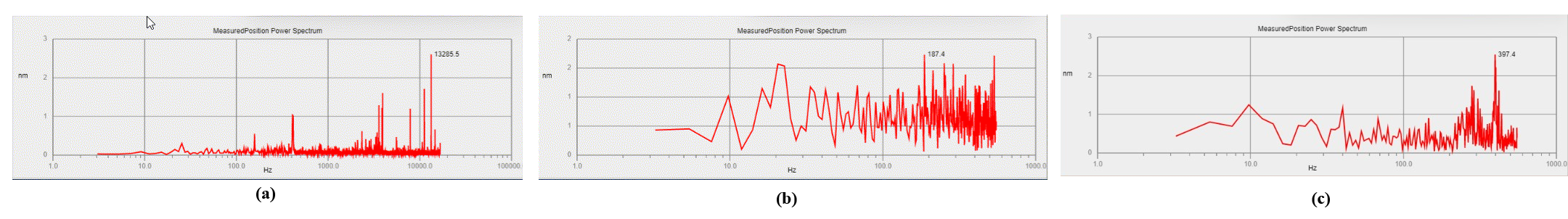}
  \caption{Examples of actuator power spectra. Illustration (a) is an example of a healthy actuator's spectrum. (b) shows a Type I oscillation spectrum. (c) shows a Type II spectrum.
  \vspace{5mm}
  \label{fig:spectra}}
\end{figure*}

The power spectrum of a healthy actuator looks like illustration (a) in Fig.~\ref{fig:spectra}. When an actuator begins to oscillate, its power spectrum takes on one of two different forms. The power spectrum shown in (b) is what we call a type I, and the power spectrum in (c) is a type II.

The difference between the two depends on the position of the actuator. In the Type I case, the actuator is able to oscillate freely. This is because there are no physical constraints on the actuator that would act to dampen its oscillation. Actuators in the two middle rings of the torus, being largely unconstrained, typically show their entire power spectrum exhibiting oscillatory behavior, what we have termed \textit{ringing}.

Actuators that are constrained by their physical conditions exhibit Type II spectra. These conditions can take several forms, the most common of which are either the boundary conditions of the actuator's placement, or its position next to an actuator that is out of service.

Boundary conditions are imposed by the actuator's ring placement. The central ring of the ASM is next to the central restraining ring of the thin shell; the motion of the mirror is constrained by the physical edge of the mirror surface, thereby constraining the motion of the actuator. Through trial and error, we have found that actuators in rings one through four (counting outwards) are constrained in this way. While tuning the actuators in these regions, we look for a response where the power spectrum shows oscillatory behavior in a small portion of its spectrum. Typically this occurs in the 2 kHz - 5kHz region of the spectrum. A constrained actuator typically begins to oscillate at this frequency, an effect we have labelled as the \textit{2k Forest}.

Actuators located in close proximity to an actuator that has been taken out of service also exhibit Type II oscillatory behavior. This position imposes a constraint to motion similar to the mirror's boundaries, in that it restricts the actuator's ability to freely oscillate. Therefore, actuators in the typically unconstrained regions may also exhibit power spectra of this type. Unfortunately, these effects are additive... An actuator located in the inner rings that is also located next to an out-of-service actuator may not be able to freely oscillate at all. We have termed this type of actuator \textit{`non-responsive'}.

Actuator Power spectra are promising in another way, aside from their use in tuning. When an actuator begins to throw wildcoil events, or begins to exhibit positional issues, its power spectrum changes. We are currently examining these phenomena to create a system for determining when an actuator might be electrically failing, loosening in its borehole mount, or even the degree to which it is being restrained by the proximal effect.

\subsection{Noise}
\label{sec:noise}

From the earliest periods of testing in the ASM laboratory, to the most current engineering runs, unexpected levels of actuator noise have been noted during mirror operation. 

The ASM was designed with a specified noise level of $<10$ nms RMS, which puts it in the same noise specification range as the LBT adaptive secondary. But at no time, until very recently, has the ASM performed with a noise level less than 10 nms RMS, and is typically more in the range of 20-35 nms RMS, with some actuators exceeding 75 nms RMS. This is not ideal behavior, and finding the cause of this noise has become the top ranked priority in troubleshooting.

\subsubsection{Actuator noise defined}

We define an actuator to be noisy when the RMS value of multiple positional measurements, taken over a brief period of time, while the actuator is commanded to hold position, exceeds 10 nm. 

This is an extremely lenient definition. For instance, the actual measured noise for the LBT ASM is around 3 - 4 nm. The measured noise for the MAPS ASM  varies wildly, from around 10 nm to as high as 50 nm. Under these conditions, of course, the very concept of positioning an actuator becomes almost meaningless, and wavefront quality degrades. 

\begin{figure}[b]
  \plotone{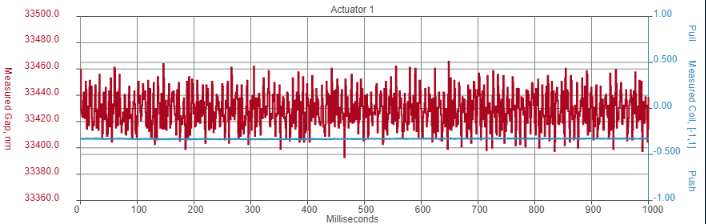}
  \caption{A Position vs Time plot for a typical actuator.
  \label{fig:pvt}}
\end{figure}

Actuator noise is essentially actuator jitter. An actuator doesn't really ever achieve a position and stay precisely there. It is constantly in motion. This can be seen in the actuator's position versus time plot, as shown in Fig.~\ref{fig:pvt}. This constant motion is why the position plot shows a collection of tightly space vertical lines... each line represents a shift in position. This, under low noise circumstances, more than likely results from errors in the capacitive sensing system, providing constant updates in measured position which the control system then tries to move the actuator to. Currently, however, most of the shift in position is due to noise.

Jitter it is a measurable and definable effect. It is also a serious problem, as the randomness it interjects into the actuator's position makes it extremely difficult for it to be commanded to a position; compounding this, it also is sometimes mistaken by the system control software as wildcoil events.

\begin{figure*}[t!]
  \center
  \includegraphics[height=385 pt]{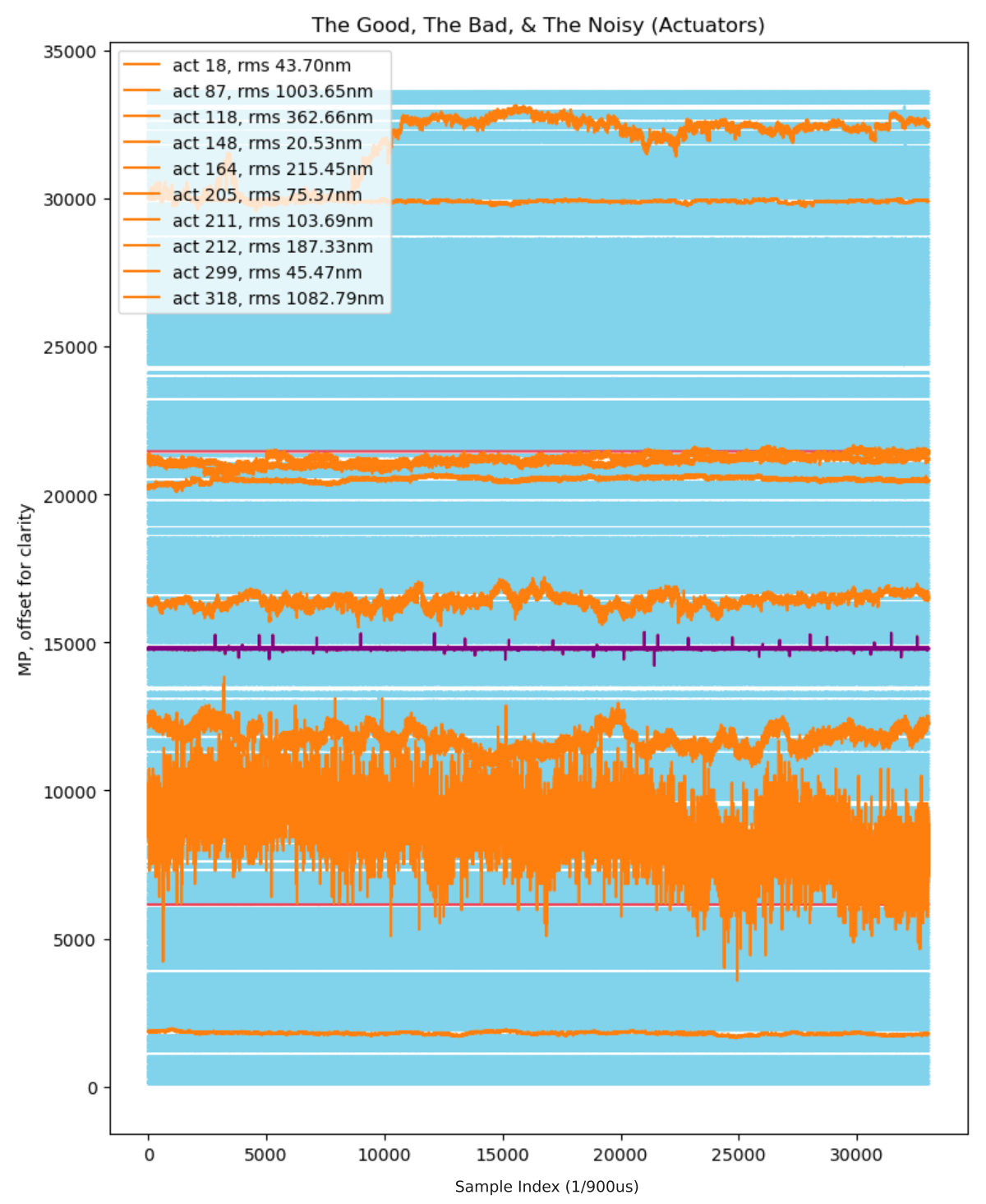} 
  \caption{A gallery of some ways ``actuator noise" can manifest. Each trace in the figure is a plot of a single actuator's reported Measured Position as sampled every 900$\mu$s, recorded by a 30-second block of continuous actuator telemetry, during which the shell was commanded to hold a static position. Blue traces represent normal actuator behavior: when commanded to stay in place, they stay in place. Orange traces, instead, show the ``wandering actuator" type of noise: when commanded to stay in place, they display varying amounts of jitter about the commanded point, often with significant excursions from where they are supposed to be. The actuator shown in purple holds its position, but has strikingly regular and periodic single-measurement excursions, perhaps indicating an electronics issue rather than physical motion.
  \label{fig:ActNoiseGallery}}
  \vspace{3mm}
\end{figure*}
   
\subsubsection{Noise measurement}

We measure ASM actuator noise by running a script that commands a mirror position and then queries the actuators via the control software INDI system. The data we retrieve (noise telemetry) is commanded position, reported position and coil current. It samples every actuator at 1.1 kHz over a set period of time, and then takes the RMS of all those measurements. The result shows, on average, how far the actuator has strayed from being stationary. This noise telemetry data showed us some very interesting thing about the ASM.

\subsection{Noisy Actuator Behavior}

\begin{figure*}[t!]
  \plotone{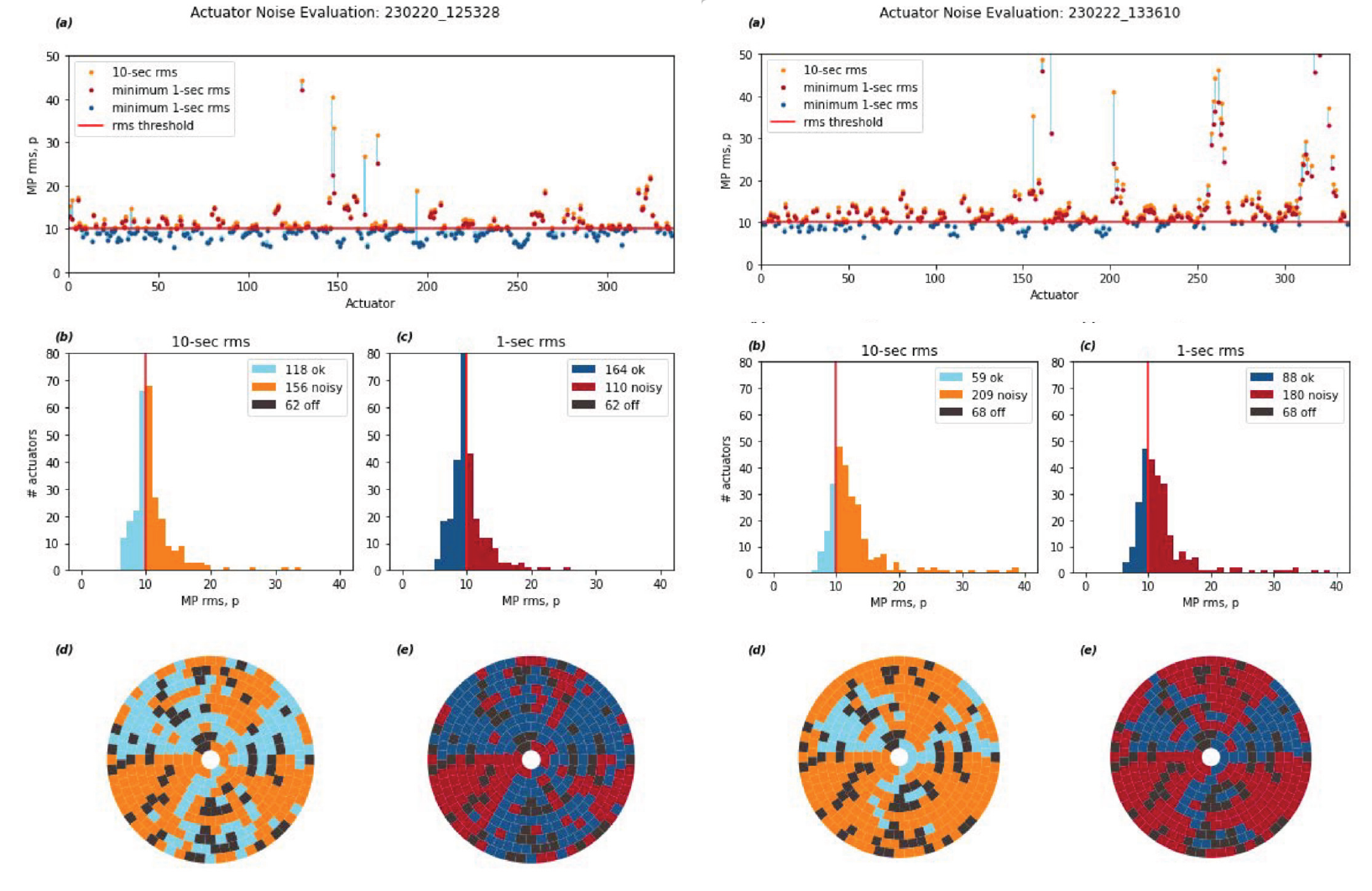} 
  \caption{Noise telemetry reports for the MAPS ASM. The report on the left shows the first telemetry report run as part of our noise investigation, the second shows the significant change brought about by unplugging and re-plugging electrical connectors that join the daughterboards to the motherboard. 
  \label{fig:noiseRep01}}
  \vspace{5mm}
\end{figure*}

Fig.~\ref{fig:ActNoiseGallery} shows multiple ways in which noise can affect an actuator. Each trace represents a single actuator's measured position, sampled every $900\mu s$ over a thrity second period.

\subsubsection{Varying noise levels.}

One of the first things we discovered was the extreme variability of measured noise. The two major factors influencing noise have now been identified as: 1) where the ASM is located (or, more precisely, to what power distribution system it is attached too), and 2) The angle of the ASM relative to the gravity vector. The telemetry data that shows the lowest noise is always, in any situation, that taken while the ASM is mounted upright with the mirror facing downward.

Our attempt to understand the excessive noise levels began with taking frequent noise telemetry. The first time we ran the script, the ASM was positioned in `zenith position' (i.e., upright, thin shell facing downward, with the mirror surface normal to the gravity vector), and was plugged into clean laboratory power. The results are shown in the illustration on the left of Fig.~\ref{fig:noiseRep01}. We took a ten second sample of actuator readouts. We then took RMS position over the course of the entire ten second sample, and did the same for a smaller one second sample. Section (a) of the report shows these two groups plotted. The red line is the 10 nm RMS cutoff for an actuator to be determined as noisy. Sections (b) and (c) show the two groups individually, and sections (d) and (e) show the position of the noisy actuators on the ASM actuator grid map.

Because of the pronounced correlation to two of the daughterboards (each daughterboard connects to a pie sliced wedge of the ASM actuator grid) we unplugged and replugged the daughterboard to motherboard connectors. This only made the situation worse, as another set of telemetry data shows, in the illustration on the right of Fig.~\ref{fig:noiseRep01}.

\subsubsection{Power and elevation}

\begin{figure}[!t]
  \plotone{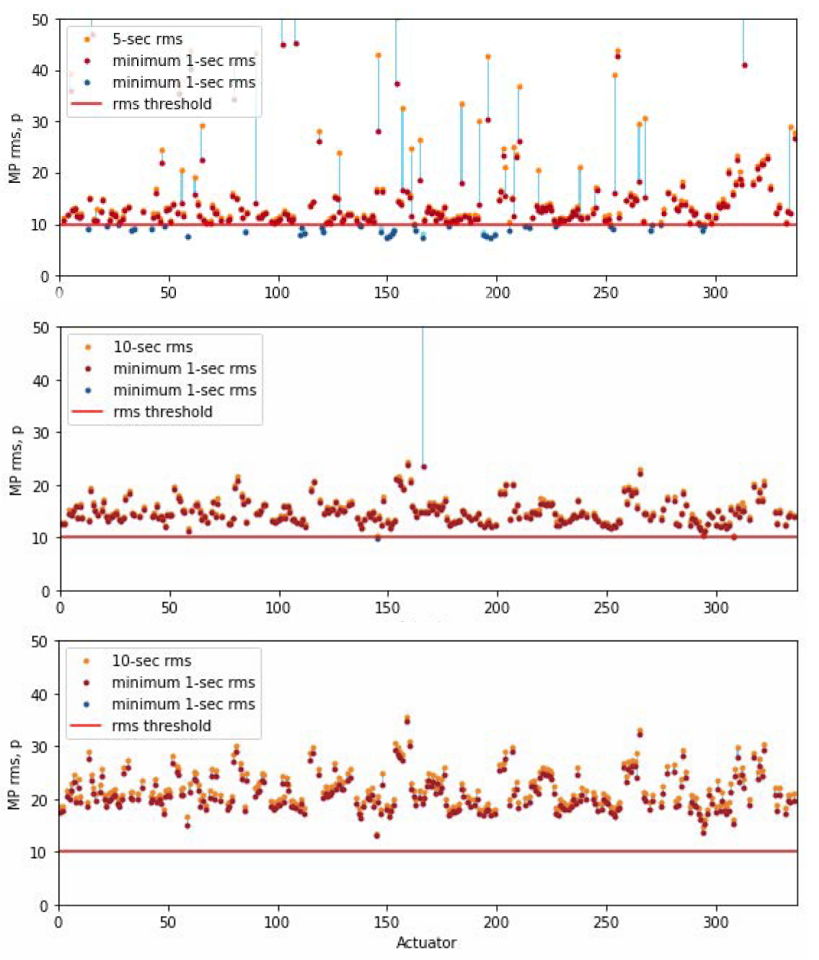}
  \caption{Noise telemetry for the MAPS ASM. The top plot is for the ASM at telescope zenith. The middle plot is for the telescope at 80 degree elevation, the bottom plot is for the telescope at 70 degrees elevation.
  \label{fig:NoiseRep02}}
\end{figure}

Noise telemetry taken during the fourth engineering run showed an dependence of noise on the telescope's elevation. This is shown in Fig.~\ref{fig:NoiseRep02}. From the top to the bottom image, these data show noise levels with the telescope at zenith, $80\degree$ elevation, and $70\degree$ elevation. 

Repeated noise level experiments further showed a dependence on the power supply the system was plugged into. The lowest noise results were consistently found with the ASM plugged into the ASM laboratory's clean power outlets, and the worst when the ASM was mounted on the telescope. The power at the MMT telescope is known to have ground loop issues, which we suspect causes high actuator noise.

\subsubsection{The tin foil hat solution.}
\label{sec:tinhat}

\begin{figure}
  \plotone{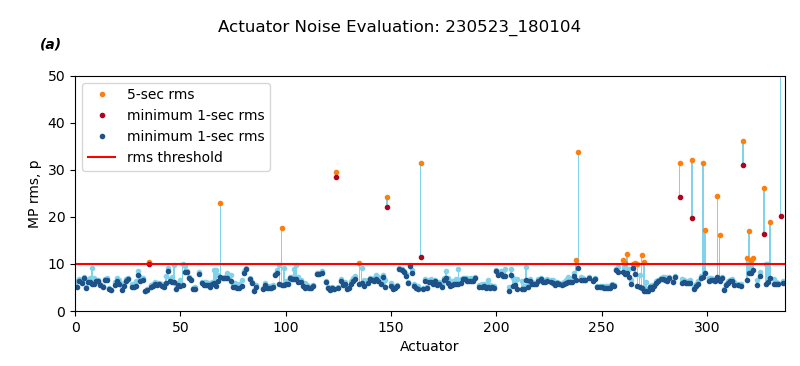}
  \caption{Noise telemetry resulting from applying the `tin foil hat solution' to the noisiest actuator. This is the lowest level of noise recorded for the MAPS ASM, and is well within design specifications. The pink line indicates the 10 nm RMS noise threshold.
  \label{fig:TinFoilRep}}
  \vspace{4mm}
\end{figure}

An electrical engineer looking into the noise issue before the ASM was dis-assembled after the fourth engineering run decided to conduct an experiment. He took a piece of tin foil and wrapped it around the top of the noisiest actuator in the ASM. Not expecting that this would do anything, but running out of ideas, we ran noise telemetry again. We were surprised, to say the least, at the result which is shown in Fig.~\Ref{fig:TinFoilRep}.

This is the cleanest noise profile that the MAPS ASM has ever produced, and it shows that the ASM can clearly maintain a level of noise at or below its specifications. The question became, why did placing a piece of tin foil on a single actuator quiet the entire ASM? To follow up, we removed the tinfoil and ran telemetry again. The noise level remained the same. 

Clearly, the foil had nothing to do with cleaning up the noise; there was a loose component, cable or connector that had been jarred in the process of applying the foil, but we were not able to identify it. 

A few days later, after the ASM has been moved, the effect vanished and the noise returned. However, the noise level was substantially reduced after the ASM was rebuilt and actuators replaced. Unfortunately, this minimum noise state has not been repeated.

\subsubsection{The problematic nature of noise}

What, exactly, does actuator noise imply for the ASM? The biggest effect of noise is that it makes it difficult if not impossible for an actuator to achieve a correct position and hold it, simply because the actuator is jittering between errant positional values. When trying to make minute corrections to phase, the mirror surface often must move in small increments of nanometers. If a few actuators are experiencing reasonable amounts of jitter, the majority of the shape can still be formed. If a majority of actuators are experiencing jitter on scales 5 to 10 times the desired change in mirror position, the noise literally drowns out the signal.

But there are two other significant concerns related to noise. The first is that the control system sees any actuator that draws current rapidly and in oscillating motion -- as an actuator will do when its position sensor tells it that it is not where it should be (it jitters up, current flow reverses to pull it down, it jitters down, and the current reverses again) -- as a wildcoil. If an actuator wildcoils enough, the mirror will safe itself shutting everything down. (Or the operator will, panicking at the onslaught of multiple wildcoil errors.)

A second concern is that, at a jitter of 20nms and higher, the force the actuator has to put on the mirror and the distance it has to move it against the mirror's influence function begins to draw excessive current. This can result in overheating and shutdown.

\subsection{Actuator Failure}
 \label{sec:failure}

The MAPS project has had a long history of actuator loss, starting from the moment it entered the laboratory after assembly. Investigating the reason actuators die, or play possum, is an important step in bringing the ASM to the level of a fully functional instrument.

\subsubsection{A taxonomy of actuator failure}
 \label{sec:taxonomy}

There are many ways that MAPS actuators cease to function. Here are the most common. (Also see Table~\ref{tab:ActuatorFailures}).

\textbf{Electrical Failure:}
A few, but by no means all, of the actuators on the database 'morgue' list have been true DEADFEEDS, electrically dead and non-responsive. We can assume that these are the result of component failure, which is an accepted risk in electronics manufacturing, and simply require replacement.

The rest of those originally presumed to be electrically dead turned out to simply have loose cables, or disconnected connectors. Some of those loose cables or disconnected connectors came from attempts to reach other cables to check their connections. Cable is forced, by sheer quantity, into a small tight space. And USB-C cables in bunches become rigid and inflexible, which makes the rings of actuators harder to access as they approach the center. Slight nudges with a hand are enough to disconnect a cable. Cables get crimped into a position that puts pressure on the actuator. The constantly shifting gravity vector on the telescope resulting from its motion yanks and pulls on the cables. Cables can be accidentally and easily reversed when replacing. Identifying labels can peel off. And cable is heavy. All of these are issues concerning the cramped space in which the USB cabling must occupy.

\textbf{Heat:}
Besides that, though, there is a bigger issue caused by the this 'cable jungle', and that is the ability of the ASM to cool itself. One of the most innovative portions of the overall design is the use of a passive, air-cooled heat dispersion system, which relies on a form of heat pipe called a heatsink pipe (see Sec.~\ref{sec:thermaldesign}) These are highly effective at rapidly transferring heat, \textit{if} there is room for air to convect it away. That fact that some actuators overheat indicates that the cabling that dominates the area on top of the cold plate is blocking airflow. Heat will eventually destroy electronics, and at a minimum, an overheating actuator’s H-Bridge (the component in the actuator circuitry that is responsible  for directing current to the voice coil) will trip its own circuit breaker and shut the actuator down. At some point after it cools, it will attempt to turn itself on again. This is the equivalent of an actuator playing possum. The solution to this problem is most likely using small fans to direct airflow through the cabling.

\textbf{Position Errors:}
By far the largest number of actuators removed from duty have been due to one form or another of error resulting from failures of their capacitive sensing system, the system that tells the actuator the vertical position of the mirror suspended in the magnetic field directly above its coil. 

Because an actuator’s two jobs are determining position and moving the mirror, and one follows from the other, loss of knowledge of position means the actuator can’t effectively move the mirror to a desired position. One way this happens is with a ‘Cap Sensor Error’, most likely the result of a failure in actuator electronics. The actuator does not report location (a red flat line on a position versus time chart), and therefore the actuator does not send power to its coil (a blue flat line on the same chart). Note that this is not DEADFEED, however, as it still reports that it is an actuator. It simply doesn’t report back position information.

Another way this happens is that the actuator doesn’t issue a particular error, like a DEADFEED or Cap Sensor Error, but triggers frequent minor warnings about exceeding its distance limits. These will cause failed flats, frequent error messages, and the occasional mirror shutdown. This eventually ends up in the actuator being removed from service by the operator. The actuator is electrically fine, apparently, happily reporting incorrect distances and the fact that it is electrically alive, but it is misbehaving and driving both its neighbors and the operator nuts.

\subsubsection{The causes of actuator error and failure}
 \label{sec:causes}

There seems to be two primary causes of actuator errors: misalignment of the actuator in the borehole, and errors in the capacitive sensing circuit.

\textbf{First Cause}
The first cause is the misalignment of the actuator in the borehole. There is a correlation between actuator failure (including those that have been disabled) and the actuator bore placement measurements taken after the mirror was disassembled. Actuators positioned too high in the hole either touched or were within one or two millimeters of the magnet. These had all been disabled for errors such as the actuator being unable to reach desired position. Actuators positioned too low in the hole were prone to high current draw and overheating.

\begin{figure}[t!]
  \plotone{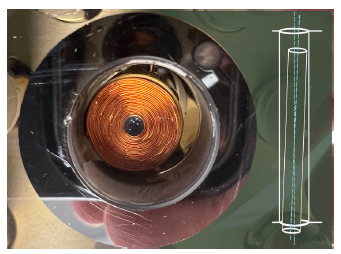}
  \caption{An actuator whose central axis is not aligned with the central axis of its borehole.
  \label{fig:alignment}}
\end{figure}

Also, many actuators were found with their central axes at an angle to the central axis of the borehole, as shown in Fig.~\ref{fig:alignment}. A coil positioned above the mirror magnet in this way does not put the expected field on the magnet. One result of this is increased power draw; another is that the field puts undesired torque on the mirror.

These actuator misalignments result from two things: a difficult-to-execute system for installing actuators, and a less-than-optimal retaining system for insuring that the actuator is maintained tightly in position after it is installed. The lack of the first leads to uncertainty in vertical placement, and the second leads to the inability to keep the actuator upright in the bore.

\textbf{Second Cause}
The second type of error comes from the coupling between the actuator and the rectangular metal surface that runs up the side of the bore. This surface joins the
chromium disc that surrounds the borehole on the top surface of the reference body and forms the second plate of the measurement capacitor. The actuator mechanically couples to this surface by pushing on a pair of pins, one on either side of the actuator, with loaded springs. The current flows through the pin into the body of the actuator where it connects to the interior electronics.

The entire success of the endeavor rides on this contact, as the voltage across it is what the actuator uses to determine position. It is, perhaps, the weakest part of the entire design. Vibration wears the metal off of the contact surface; the contact surface disconnects from the ring at the top of the hole; the tip of the pins oxidizes. A pin’s spring sticks. Any of these degradations to the mechanism causes the impedance of the circuit to change. Changing the impedance changes the voltage, and therefore the distance 
measured.

\subsection{Software: Integration and Safeguards}
 \label{sec:softwaresafety}

Each subsystem in the MAPS AO path -- telescope, ASM, WFS, and science instruments -- is controlled by its own software, written by multiple teams from different institutions, with different purposes in mind. The overall challenge of MAPS integration is to link these disparate components into a single coherent whole.

\subsubsection{Inter-software links}

The software used so far to operate the MAPS ASM can be divided into three parts: \textbf{MMTAO-Main}, \textbf{Chai}, and \textbf{Cacao}. MMTAO-Main is the original software developed for the ASM, i.e. it communicates with INDI, has web-based interfaces, and is currently used for powering the ASM, setting flats, and monitoring ASM performance and safety. Chai integrates the ASM and wavefront sensors. It communicates with gRPC and implements slope calculations for the wavefront sensors. It also forwards commands from Cacao to the ASM. Cacao~\citep{Guyon18} takes the wavefront output of Chai and calculates actuator commands. The telescope itself, including the secondary mirror hexapod, is yet another separate system, and is entirely controlled by the MMTO operator. 

In operation, the two major linkages are the \textbf{Chai} connection sending commands \textit{to} the ASM, and the \textbf{telescope control} connection receiving commands \textit{from} the mirror to offload low-order Zernike aberrations by means of optics or mount movement.

\subsubsection{ASM command safety}

The requirements are straightforward, but safe implementation among multiple pieces of software requires a firm understanding of what assumptions each side is making, where pitfalls may arise, and how best to check or fix them. During our first runs on sky, for instance, we were tripped up by misaligned units between WFS output and ASM input -- the ASM is programmed to accept position commands in nm -- and although no damage was done, it served as a forcible reminder: in the presence of easily-made errors with enormous consequences, we should not rely entirely on our lowest-level actuator safeguards. 

The original ASM software was written with only two error states. The first, analogous to the ``RIP" or ``TSS" states of LBTO/MagAO units, applies a strong current to all coils and clamps the shell against the reference body. The second, for circumstances yet more dire, simply powers the unit off. In operation, though, we would like something gentler, more like the ``skip frames" feature of LBTO/MagAO, which simply does not apply any command flagged as dangerous. In that spirit, we have decided to implement gatekeeper oversight for all ASM commands \textit{before} they are sent to the unit.

Each actuator of the ASM can accept commands as either a position setpoint in nm, a coil current setpoint in ``Elwoods" (ewd; a scaled current ranging from -1 to +1), or a combination of both in a one-two punch of feed-forward current plus new position. Accordingly, we need limits on both. 

We have so far only implemented the suite of position limits, each set of which is intended to protect against particular dangers. Absolute position limits (15,000 nm and 100,000 nm) ensure that we do not physically contact the reference body or small dust particles. Limits relative to the optical flat (+/-10,000 nm) allow for reasonable correction (typical atmospheric deviation is within +/-5,000 nm) but avoid actuator overheating caused by the high coil currents needed for large-amplitude high-spatial-frequency commands -- overheating that not only stresses the components, but loses 15-30 minutes of sky time to recovery. We use checks on shell-wide mean deviation from flat (5,000 nm) and maximum deviation from previous command (10,000 nm) as checks against abnormal behavior and physically implausible commands. We anticipate that these values will need tuning on sky, but they serve as a reasonable starting point.

We plan a similar set for coil current, with the important addition of a check on mean current across all actuators. The goal there is to avoid accidentally detaching the shell: on the one hand, any single actuator needs to be able to exert force between the full -1 and +1 ewd; on the other, sending -0.1 ewd to all actuators will completely separate the shell from the reference body, leaving it no longer under active control but simply resting on physical retaining clips. More alarmingly, one might imagine that sending -1 ewd to all actuators could eject the shell with some force. 

\subsubsection{Low-order offloading}

Aberrations of the lowest Zernike modes -- tip, tilt, focus, the first orders of comas and astigmatisms -- are, under normal atmospheric conditions, the largest in magnitude. They can quickly eat up all of the available actuator stroke on the ASM, or all of the dynamic range of the WFS. Further, they are both caused by and constantly changing in response to the more general environment of temperature and gravity, and to properties of the telescope itself. 

Common practice with adaptive secondaries is to ``offload" these aberrations. Gross and slowly varying tip, tilt, focus, and coma can all be countered by fairly straightforward hexapod moves, and astigmatism by primary mirror actuators. If these moves are fed to the telescope every handful of seconds, they can free up actuator stroke and limit overall mirror command amplitude, both of which significantly improve AO correction. Offloading also blunts the effect of the steep temperature changes so typical of desert twilights, and allows the ASM to work in the well-characterized middle of its operational range rather than the outer wilds.

We do not yet have continuous offloading systems in place for MAPS. That lack has been a significant hurdle in observing runs to date, and we hope to begin limited implementation soon.

\section{FUTURE WORK}
 \label{sec:future}

We have demonstrated the bare-bones functional aspects of the MAPS ASM, now on sky as part of a complete AO system. Our challenge going forward is to finish the job -- to work steadily towards an ASM that will be an effective and reliable component of our quest for great science. 

Our most immediate concerns are (a) actuator noise, and (b) command safety. Having now established that there is no intrinsic fault in the components themselves, we plan to proceed with more systematic measures against ground loops, electronic interference, and the like. We expect anti-noise work to be an ongoing process, but it can proceed alongside more directed activities. On the safety front, we are in the process of implementing a more complete set of software checks across all commands sent to the mirror, especially prioritizing checks on feed-forward currents, where simple mistakes could present a significant risk to the shell. 

Once we are confident of safety and can reliably control noise, we will tackle some of the factors limiting correction quality. The proximal effect is one such: what's the best way to effectively float the shell over disabled actuators in the presence of bias magnets? Another is actuator `jumps', or sudden and temporary uncommanded single-actuator motion, perhaps also linked to the proximity problem. 

A separate major push involves the optical test stand, improvements to which will allow us off-sky optical confirmation of calibrations and commands. We will evaluate strategies for generating a better off-sky optical flat, with particular focus on the outermost rings of actuators.

Our eventual goal is to confidently meet ASM requirements as defined by MAPS science: we will correct at least 160 but up to 300 modes, settling within 1ms, and using actuators which do not themselves add significant wavefront error. We have much work to be done, but the ability and opportunity to integrate the ASM with the rest of MAPS, and to use the unit on-sky, will only help us: we gain immense insight about our mirror from watching it interact with the other AO components. We can't, given the geometry of our telescope, make an artificial laboratory star, but the natural world awaits us, and, as our observing runs continue, we will make good use of it.

\section{Acknowledgments}
This is a preprint version of SPIE article OP432-53, to appear in the Proceedings of the SPIE Unconventional Imaging, Sensing and Adaptive Optics 2023 Conference.

The MAPS project is primarily funded through the NSF Mid-Scale Innovations Program, programs AST-1636647 and AST-1836008. This research has made use of NASA's Astrophysics Data System. We respectfully acknowledge the University of Arizona is on the land and territories of Indigenous peoples. Today, Arizona is home to 22 federally recognized tribes, with Tucson being home to the O’odham and the Yaqui. Committed to diversity and inclusion, the University strives to build sustainable relationships with sovereign Native Nations and Indigenous communities through education offerings, partnerships, and community service.)

\bibliography{report.bib}

\end{document}